\documentclass[12pt,preprint]{aastex}
\usepackage{natbib}
\begin{document}

\title{Constraining the Evolutionary Stage of Class I Protostars:
Multi-wavelength Observations and Modeling}

\author{J.A. Eisner, L.A. Hillenbrand, and John M. Carpenter}
\affil{California Institute of Technology, Department of Astronomy, MC 105-24,
Pasadena, California 91125}
\email{jae, lah, jmc@astro.caltech.edu}

\and
\author{S. Wolf}
\affil{Max-Planck-Institut f\"{u}r Astronomie, K\"{o}nigstuhl 17, D-69117,
Heidelberg, Germany}
\email{swolf@mpia-hd.mpg.de}




\begin{abstract}
We present new Keck images at 0.9 $\mu$m and OVRO 1.3 mm continuum
images of five Class I protostars in the Taurus star forming 
region. We analyze these data in
conjunction with broadband spectral energy distributions and 8-13 $\mu$m
spectra from the literature using a Monte Carlo radiative transfer code.
By fitting models for the circumstellar dust distributions
simultaneously to the scattered light images, millimeter
continuum data, and the SEDs, we attempt to distinguish between flared disks,
infalling envelopes with outflow cavities, and combinations of disks
and envelopes.  For each of these circumstellar density distributions,
we generate grids of models for varying geometries, dust
masses, and accretion rates, and determine the best fits by minimizing the
residuals between model and data.
Comparison of the residuals for best-fit disk, envelope, and disk+envelope 
models demonstrates that in general, models incorporating {\it both}
massive envelopes and massive embedded disks fit the imaging+SED data best.  
The implied envelope infall rates for these disk+envelope models are
generally consistent with infall rates derived by previous investigators,
although they are approximately an order of magnitude larger than inner
disk accretion rates inferred from recent spectroscopic measurements.
In addition, the disk masses inferred from our models are close to or larger
than the limit for gravitationally stable disks, indicating that Class I
disks may undergo periodic episodes of enhanced accretion, perhaps as
a result of gravitational instabilities.  An important caveat to these
results is that in some cases, no single model can fit all of the imaging and 
SED data well, suggesting that further refinements to models of the
circumstellar dust distributions around Class I sources are necessary.
We discuss several potential improvements to the models, as well as new
constraints that will become available
with upcoming millimeter  and infrared facilities.
\end{abstract}

\section{INTRODUCTION \label{sec:intro}}
The canonical picture of low-mass star formation is that of a rotating,
collapsing cloud of dust and gas that forms a protostar surrounded
by a disk \citep[e.g.,][]{TSC84,SAL87,SHU+93}.  Different stages of
this theoretical evolutionary process have been equated with observed
differences in the spectral energy distributions
(SEDs) of protostellar objects, which have been grouped
into classes 0-III based on their infrared spectral index and
the ratio of sub-millimeter to bolometric luminosity \citep{LW84,LADA87,ALS87,
AWB93}.   
In this classification scheme,
Class 0 and I sources are still in the main accretion phase, and emit
most of their radiation at far-IR and submm wavelengths due to reprocessing
of light from the central protostars by dust grains in an infalling envelope.
In contrast, Class II and III sources exhibit directly revealed 
pre-main-sequence stars in addition to emission from circumstellar disks.

This classification scheme is an attempt to represent discretely a  
continuous evolutionary sequence, and there are transition objects that create 
some blur between classes.  
Moreover, this sequence is defined solely from spatially
unresolved SEDs, which contain only limited
information about the circumstellar distributions.  Thus, it is not clear
that the observed differences in SEDs truly correspond to evolutionary
changes in the circumstellar geometry.  A crucial test is to constrain 
the geometry of material around members of these different evolutionary 
classes using spatially resolved images, and
thereby either confirm or refute the evolutionary sequence inferred from
spatially unresolved SEDs.

Direct imaging of Class II sources has shown that the bulk of the
circumstellar material lies in disks 
\citep[e.g.,][]{MO96,KS95,EISNER+03,EISNER+04}, while there may be a small 
amount of material in tenuous envelopes \citep[e.g.,][]{GRADY+99,SEMENOV+04}. 
Thus, Class II sources appear to be fully-assembled
young stars surrounded by rotating disks from which they continue to 
accrete material.  
Direct images of the less evolved Class I sources are relatively 
rare, due to the large extinctions to these embedded objects, and thus
the circumstellar geometry for these objects is not well constrained.

Studies of the emergent spectral energy distributions at wavelengths 
$\ga 10$ $\mu$m have provided important, albeit ambiguous, constraints on the 
circumstellar dust distributions for Class I objects.  Models incorporating
infalling, rotating, envelopes with mass accretion rates on the order
of $10^{-6}$ M$_{\odot}$ yr$^{-1}$ are consistent with observed SEDs 
\citep[e.g.,][]{ALS87,KCH93}, and the compatibility of the  
derived mass accretion rates with statistically-inferred ages of
Class I sources \citep{ALS87,MYERS+87,BM89,KENYON+90} provides further support
for these envelope models.
However, for some objects
whose SEDs can be explained by spherically-symmetric dust
distributions, it has been suggested that nearly edge-on flared disk models 
may also be able to reproduce the observed SEDs \citep{CG99}.  
Observations that spatially resolve the circumstellar emission are necessary
to remove the ambiguities inherent in SED-only modeling.

The circumstellar geometry is constrained directly by spatially resolved
images.  Scattered light at near-IR wavelengths 
\citep[e.g.,][]{KENYON+93,WKG97,WHITNEY+03a}, with asymmetric emission 
morphologies observed in some sources, provides strong evidence that the
circumstellar material is not spherically distributed around Class I objects.
However, since scattered light traces tenuous dust in the surface layers
of circumstellar dust distributions, it does not allow an unambiguous
determination of geometry or other circumstellar properties.  For example,
while spherically-symmetric dust distributions can be ruled out, 
we show below that modeling
of scattered light alone can not necessarily 
distinguish between a flattened envelope and a flared disk.

Modeling of images at multiple wavelengths can provide tighter constraints
on geometry, since 
emission at different wavelengths arises in different layers of the 
circumstellar material, with short-wavelength scattered light from
low-density surface layers and longer-wavelength emission from
deeper, cooler layers at larger radii.
For the Class I source L1551 IRS 5, spatially resolved observations 
at several wavelengths 
\citep[e.g.,][]{SSV76,KM90,BUTNER+91,LAY+94,LADD+95,RODRIGUEZ+98,CR00,MA01} 
and detailed spectroscopy \citep{WHITE+00} have been
combined with SED modeling, providing important additional constraints
on the models \citep{OSORIO+03}.  
Moreover, the combination of low and
high resolution millimeter observations facilitated distinction of compact
disk emission and more extended envelope emission.  Similar modeling of
multi-wavelength observations for IRAS 04302+2247 provided firm
constraints on the distribution of circumstellar material and the
properties of the circumstellar dust grains \citep{WPS03}.

To further our understanding of the Class I population as a whole,
we have obtained high spatial resolution observations of five additional
Class I sources
at multiple wavelengths which, when combined with SEDs, enable much tighter
constraints on circumstellar dust models than available from any single 
dataset.  We present images of scattered light at 0.9 $\mu$m
and thermal emission from dust at 1.3 mm.  In addition,
we augment SEDs compiled from the literature with new photometry at 0.9 $\mu$m,
18 $\mu$m, 1.3 mm, and 3 mm
wavelengths.  Using the three dimensional radiative transfer code MC3D
\citep{WH00,WOLF03},
we model our data in the context of three types of circumstellar dust
distributions: 1) rotating infalling envelopes; 2) flared disks; and
3) combinations of envelopes+disks.   

The best fits are obtained for models incorporating both an
envelope and an embedded disk, although we show that pure disks or envelopes
can reproduce certain aspects of our data.  For each source, 
we discuss the properties of the best-fit models, including the geometry of
the dust distributions and implied mass accretion rates.  
We use these results to help place Class I sources in the proper
evolutionary context.  Finally, we discuss refinements to the
models that may improve agreement with observed multi-wavelength data, 
and describe new constraints that will become available from 
upcoming millimeter and infrared facilities.

\section{OBSERVATIONS \label{sec:obs}}

\subsection{The Sample \label{sec:sample}}
Our sample consists of five Class I sources in the Taurus star forming region
drawn from the larger sample of protostars studied by \citet{KCH93}.
This subset was selected based on three criteria: 1) sources must be detected
at red optical wavelengths ($F_{\rm 0.9 \mu m} \ga 10^{-5}$ Jy), 2) this emission
must appear resolved, and 3) sources must 
emit strongly at millimeter wavelengths ($F_{\rm 1mm} \ga 10^{-2}$ Jy). 
We initially observed the complete \citet{KCH93} sample at 0.9 $\mu$m with
Keck/LRIS (see \S \ref{sec:lris} and Appendix \ref{sec:app_lris}), and
chose only objects that satisfy our selection criteria for high
angular resolution millimeter observations (\S \ref{sec:ovro}) and detailed 
modeling.  Our sample includes IRAS 04016+2610, IRAS 04108+2803B, IRAS
04239+2436, IRAS 04295+2251, and IRAS 04381+2540.
General properties of our sample are discussed here, and
information about individual objects is given below in \S \ref{sec:results2}.

The bolometric luminosities of our sample span 0.4 to 3.7 L$_{\odot}$,
similar to the distribution of luminosities for the complete sample
of \citet{KCH93}, which has a median luminosity of 0.7 L$_{\odot}$.  The
SEDs for our sample objects are similar in shape, exhibiting high
extinction of the central objects, small amounts of mid-IR absorption,
and peaks near 100 $\mu$m.  While these SED characteristics are common
to most other sources in the \citet{KCH93} sample, several of the SEDs
for their sample exhibit deep 10 $\mu$m absorption features.  As discussed
below, the depth of the 10 $\mu$m feature is correlated with source
viewing angle; the dearth of objects with deep 10 $\mu$m absorption
in our sample may therefore indicate selection against edge-on sources.

There are other clear selection effects for our sample, since we are
biased toward sources exhibiting bright scattered light
and millimeter continuum emission.   For example, 
by selecting sources exhibiting bright scattered light nebulae, we are 
selecting against face-on sources where the directly visible, point-like
protostar would dominate the short-wavelength emission, as well as objects with
extremely high line-of-sight extinctions, such as one might expect for
massive spherical envelopes. Since edge-on sources produce dimmer scattered
nebulosity, these may also be under-represented in our sample.  Finally,
our selection criteria for bright millimeter emission favors objects with 
higher masses of dust concentrated in smaller volumes.  Future observations
of larger samples with more sensitive instruments are necessary to investigate
these potential biases.

\subsection{OVRO Observations \label{sec:ovro}}
We imaged each source in our sample at both 1.3 mm (230 GHz) and 
3 mm (115 GHz) with the
OVRO Millimeter Array between 2002 August and December.  The 3 mm observations
were conducted in the ``compact'' array configuration, which provides 
baselines between 20 and 55 m, while the 1.3 
mm observations were obtained mostly 
with the ``equatorial'' configuration, providing baselines from 
30 to 120 m.  For IRAS  04295+2251 and IRAS 04381+2540, 1.3 mm
data were also obtained
with the ``high'' configuration, which provides baselines from 35 to 240 m.
Continuum data were recorded in four 1-GHz channels.  
We calibrated the 
amplitudes and phases of the data using quasars near on the sky 
($\la 20^{\circ}$) to our target sources.  The flux of our target sources
was calibrated using observations of Neptune and Uranus, which have known
millimeter fluxes, to calibrate the response of the instrument.
All data calibrations were performed using the OVRO software 
package MMA \citep{SCOVILLE+93}.  Using the calibrated data for each
source, we inverted the interferometric visibilities to create an image,
and de-convolved and CLEANed the images using the MIRIAD package \citep{STW95}.
We averaged the data using robust weighting (with a robustness parameter of 
0.5) to obtain a good balance between sensitivity and angular resolution. 

\begin{figure}
\plotone{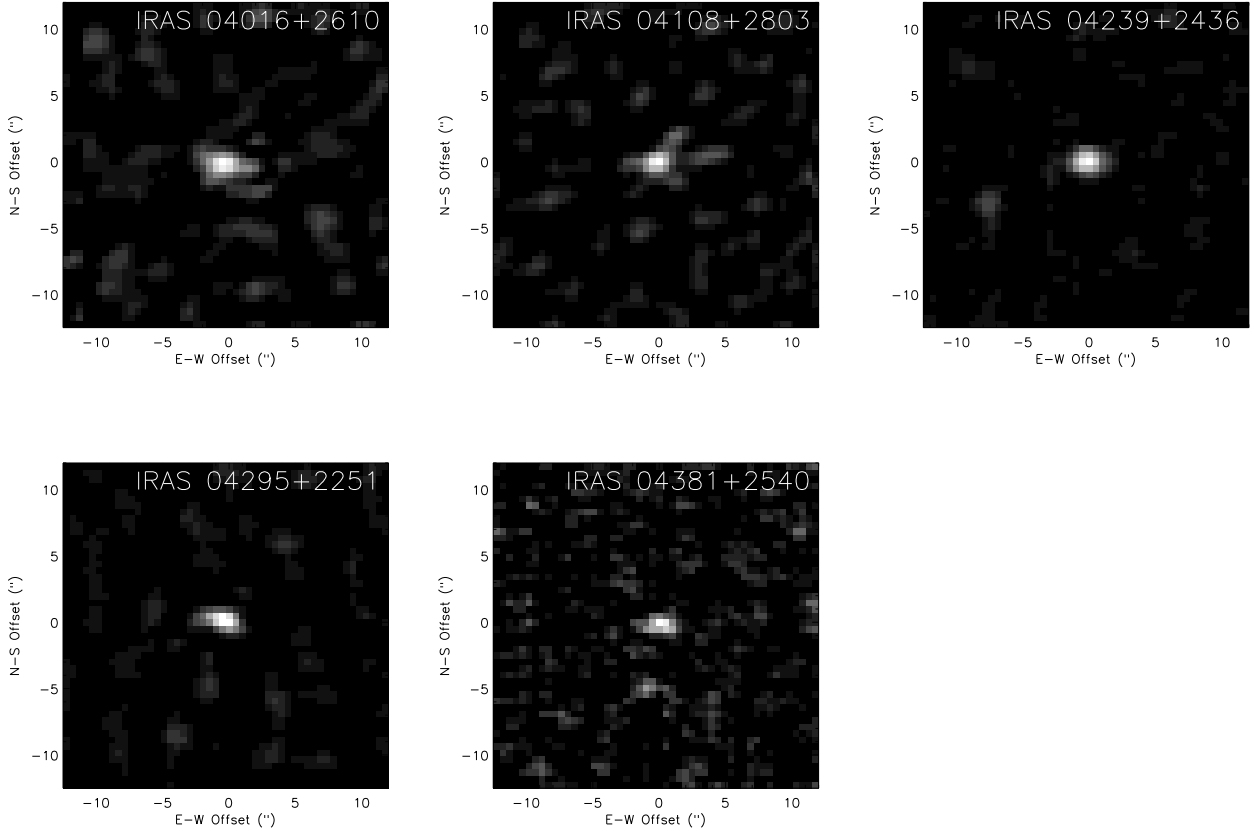}
\caption[$\lambda$1.3 mm images of Class I sample]
{Images of our sample in thermal emission at 1.3 mm wavelength,
obtained with OVRO.  The origin of each image corresponds to the
centroid of the millimeter emission.  The J2000 coordinates at the
origin of the images are (04$^{\rm h}04^{\rm m}43\rlap{.}^{\rm s}11$,
+26$^{\circ}18'56\rlap{.}''5$) for IRAS 04016+2610,  
(04$^{\rm h}13^{\rm m}54\rlap{.}^{\rm s}72$, +28$^{\circ}11'33\rlap{.}''0$)
for IRAS 04108+2803B,  (04$^{\rm h}26^{\rm m}56\rlap{.}^{\rm s}29$,
+24$^{\circ}43'35\rlap{.}''1$)  for IRAS 04239+2436,
(04$^{\rm h}32^{\rm m}32\rlap{.}^{\rm s}06$, +22$^{\circ}57'26\rlap{.}''3$)
for IRAS 04295+2251, and 
(04$^{\rm h}41^{\rm m}12\rlap{.}^{\rm s}71$, +25$^{\circ}46'35\rlap{.}''3$)
for IRAS 04381+2540.
\label{fig:mm}}
\end{figure}

From our data, we measured photometric fluxes for our targets at both 3 mm 
and 1 mm wavelengths.  The emission is much stronger at 1 mm, as expected
for thermally-emitting dust, and thus the sources
are more clearly detected at the shorter wavelength.  In addition, since
the angular resolution is inversely proportional to the observing wavelength,
and because we obtained longer-baseline data at 1 mm than at 3 mm,
our 1 mm images have better angular resolution.  Our
1 mm images, shown in Figure \ref{fig:mm}, have angular resolution 
of $\sim 2''$, and rms sensitivity of $\sim 1.5-2.5$ mJy/beam.
The measured fluxes, listed in Table \ref{tab:seds}, are integrated
over the OVRO beam, which is $\sim 2''$ at 1 mm, and $\sim 6''$ at 3 mm.

\subsection{Keck/LRIS Observations \label{sec:lris}}
We obtained images of our Class I
sample at 0.9 $\mu$m (Cousins $I$-band) using the
W.M. Keck II telescope and the Low Resolution Imaging Spectrograph
\citep[LRIS;][]{OKE+95} in imaging mode.  The observations were conducted
on 1998 October 30-31 and 1999 December 13.  The integration time was
300 seconds, the seeing was $0\rlap{.}''5-0\rlap{.}''6$, 
the field of view was $6' \times 8'$, and the plate scale was
0.21 arcseconds per pixel.  Image processing
included bias subtraction and flat fielding.  The photometric fluxes
of the target sources, integrated over a $6\rlap{.}''3$ (30 pixels) diameter 
aperture with background measured over an annulus extending from $26''$
to $30''$ (125-145 pixels), 
were calibrated using equatorial standards from \citet{LANDOLT92},
assuming typical extinction coefficients for Mauna Kea
\citep[0.07 mag/airmass at $I$ band;][]{BBD88}; all sources
were observed at an airmass of less than 1.1.

\begin{figure}
\plotone{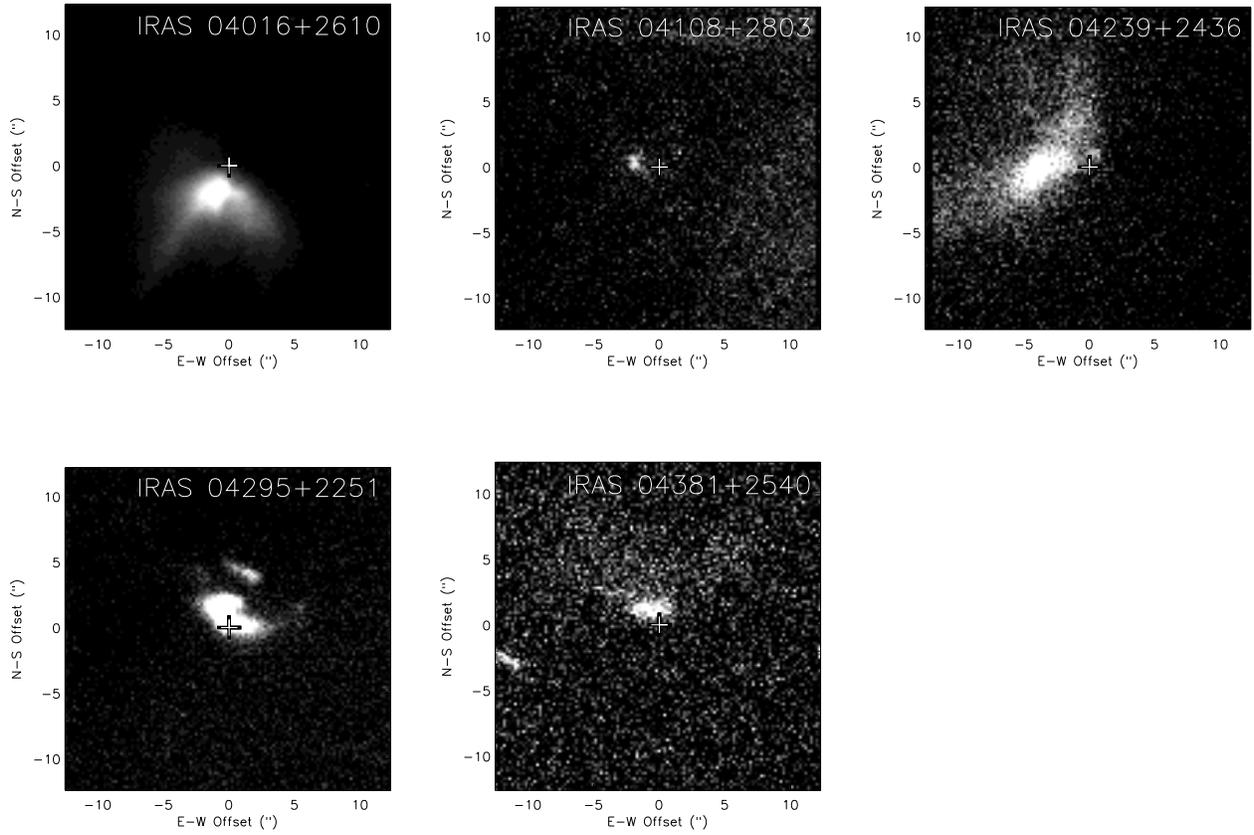}
\caption[$\lambda$0.9 $\mu$m images of Class I sample]
{Images of our sample in scattered light emission at 0.9 
$\mu$m wavelength, obtained with Keck/LRIS.  The images are registered
with respect to the centroid of the millimeter emission for each
source (Figure \ref{fig:mm}), indicated by a cross.  
\label{fig:iband}}
\end{figure}

All objects in our sample were detected at 0.9 $\mu$m, and show varying amounts
of extended emission arising from scattered light due to dust-obscured 
central sources.  Our images are shown in Figure \ref{fig:iband}.
We determined absolute astrometry for these images by finding 
reference sources in common
to our images and the 2MASS survey and computing a six coefficient fit
for the image coordinates, plate scale, and distortion.  The astrometry
is accurate to $\sim 0\rlap{.}''3$, including the uncertainty in our
astrometric solution and the overall uncertainty from 2MASS.

Because we determine accurate positions for our $I$-band images, we can 
register the scattered light emission with the millimeter-wavelength emission
observed with OVRO (\S \ref{sec:ovro}).  Since the longer wavelength
radiation is optically thin, it traces dust mass and is therefore likely
to be centered on the central source.  In Figure \ref{fig:iband},
we have indicated the centroid of the millimeter continuum emission 
(and thus the likely position of the central protostar) with a cross.
As we discuss below, the offset of the scattered light emission with respect to
the central protostar provides valuable insights into source geometry and 
viewing angle.

\subsection{Keck/LWS Observations \label{sec:lws}}
Our sources were also observed at 10.3 and 17.9 $\mu$m with the Keck 
Long Wavelength Spectrometer \citep[LWS;][]{JP93}
between 1999 August and 2000 December.  LWS allows diffraction-limited
(FWHM=$0\rlap{.}''2$) imaging over a $10''$ field of view, although the
observations described here were obtained under poor seeing conditions,
FWHM=$0\rlap{.}''3-0\rlap{.}''6$.  With
this seeing-limited resolution, none of our targets are angularly resolved at
$10-18$ $\mu$m.
Photometric fluxes were measured in a $1\rlap{.}''92$ (24 pixels) 
diameter aperture, with subtraction of the sky background measured in 
an annulus
spanning $4\rlap{.}''0$-$6\rlap{.}''4$ (50-80 pixels).  Flat-fielding did
not improve the data quality, and was therefore not applied before measuring
photometry.  Standard stars were observed each night to obtain atmospheric
extinction curves, and curve-of-growth corrections were applied to each star
to convert measured aperture photometry to infinite-aperture values.
In addition, these Keck/LWS observations provided
8-13 $\mu$m spectra, which have been reduced and analyzed by 
\citet{KESSLER-SILACCI+05}.

\subsection{Spectral Energy Distributions \label{sec:sed}}
SEDs were constructed using photometric fluxes from the literature and new
measurements at 0.9 $\mu$m (\S \ref{sec:lris}), 18 $\mu$m
(\S \ref{sec:lws}), 1.3 mm, and 3 mm (\S \ref{sec:ovro}).  
Uncertainties are not available for much of the photometry from
the literature, and therefore we
adopt estimated, uniform photometric uncertainties in the modeling
described below (\S \ref{sec:model}).  Since these uniform error bars
lead to the over-weighting of higher fluxes, 
we consider the logarithms of the SEDs in our 
modeling.  The fluxes used in our analysis, 
converted into units of Jy, are listed in 
Table \ref{tab:seds} and plotted
along with best-fit models below in \S \ref{sec:results}.

\section{MODELING \label{sec:model}}
We use the three dimensional Monte Carlo radiative transfer
code MC3D \citep{WH00,WOLF03} to model
simultaneously the SEDs, 0.9 $\mu$m scattered light images, and millimeter 
continuum images of our sample of Class I sources
\citep[MC3D has been validated by benchmark comparison to other Monte Carlo and
grid-based radiative transfer codes;][]{WHS99,PASCUCCI+04}.  
The basic model
is that of a central emission source surrounded by obscuring dust of
arbitrary geometry.
The MC3D code solves the radiative transfer equation and 
determines the temperature distribution of circumstellar dust 
self-consistently.  Moreover,
the code takes into account absorption and multiple scattering events.
The main inputs for this modeling are 1) parameters of the central star,
2) properties of dust grains, and 3) geometry of circumstellar dust.
We now describe which properties in these categories are varied in our
modeling, as well as our assumptions for other parameters.

Since our concern is primarily with understanding the geometry, we fix most
of the stellar and dust parameters. 
For the models described below, we assume that the central object resembles
a typical Class II T Tauri star.  We therefore assume
 $T_{\ast}=4000$ K, $R_{\ast}=2$ R$_{\odot}$, and $M_{\ast}=0.5$ M$_{\odot}$ 
\citep[e.g.,][]{GULLBRING+98}.
Although few direct constraints exist on these parameters,
spectroscopy of IRAS 04016+2610 has yielded an estimated effective temperature
between $\sim 3300-4200$ K and a radius of $\sim 0.9$ R$_{\odot}$
\citep{WH04,ITI04}, reasonably consistent with our assumptions.  
The dynamically-estimated stellar mass of IRAS 04381+2540 is $\sim 0.2-0.4$
M$_{\odot}$ \citep{BC99}, also compatible with our assumptions.

Accretion onto the stellar surface may generate substantial luminosity
in shocks near the stellar surface, which would supplement the stellar
luminosity \citep{CG98,GULLBRING+00}: $L_{\rm central}=L_{\ast}+L_{\rm acc}$.  
In our modeling, we initially assume that $L_{\rm central}=L_{\ast}$, 
but we include a scale factor to allow for additional accretion luminosity
(and/or slightly different stellar parameters from those assumed).  
As the last step in our modeling procedure, we find the value of
$L_{\rm central}$ for which no scale factor is required.
Although the spectral shape of emission from the accretion shock may differ
from the stellar emission, we ignore this effect and assume that the
temperature of all emission is 4000 K; the spectral shape is not critical
given the extremely high optical depths in the inner regions of our models.
In summary, all stellar parameters are constant from model to model, with
the exception of a luminosity scaling that represents $L_{\rm acc}$.

We assume that the dust grains are spherical with
a power-law size distribution $n(a) \propto a^{-3.5}$ \citep{MRN77},
where $a_{\rm min} = 0.005$ $\mu$m and $a_{\rm max}=1.0$
$\mu$m, appropriate for ISM-like grains.  
We further assume that the dust is composed of a standard ISM mixture
of 62.5\% silicate and 25\% ortho + 12.5\% para graphite 
\citep[e.g.,][]{DM93,WD01}, with optical properties from \citet{DL84}.  
Previous authors have explored the effects of varying the chemical composition
and particle size distribution of the circumstellar dust
\citep[e.g.,][]{DALESSIO+99,DCH01}.  These parameters
affect the overall SED, and 10 and 18 $\mu$m silicate features, 
to some extent.   
For simplicity, we keep the dust composition and particle
size distribution fixed in our modeling.

The main remaining input for radiative transfer
is the density distribution, and we focus our efforts on exploring the
parameter space associated with circumstellar geometry.
We consider three classes of models:
rotating infalling envelopes, flared disks, and envelopes+disks.  Each of 
these models is defined from an inner radius, $R_{\rm in}$, to an outer radius,
$R_{\rm out}$.  The inner radius is assumed to be 0.1 AU (comparable to the
dust sublimation radius), although this 
parameter is not crucial to the modeling given the high 
optical depths near to the
central protostar.  However, $R_{\rm out}$ may have substantial effects on
models for the circumstellar material, and thus we allow this parameter to
vary in our modeling.

For each of the models considered in the following sections, several 
important parameters influence the density distributions, with concomitant
effects on observed images and SEDs.
In order to explore the effects of these parameters, we generate small grids of
models by varying several important properties.  Large grids are not 
possible due to the long run-time of Monte Carlo radiative transfer codes 
(typically several hours per model), and we therefore consider only three
values for each parameter.

For the pure envelope model
(\S \ref{sec:env}), we vary the mass infall rate, $\dot{M}$,
the centrifugal radius, $R_{\rm c}$, the outer radius, $R_{\rm out}$,
and the inclination, $i$.  The values of $\dot{M}$ are chosen to provide
total envelope masses\footnote{All quoted envelope and disk masses
increment the computed dust mass by an assumed gas to dust mass ratio of 100.} 
of $5 \times 10^{-3}, 10^{-2},$ 
and $5 \times 10^{-2}$ M$_{\odot}$; since the envelope mass depends to some
extent on parameters other than $\dot{M}$ 
(see Equation \ref{eq:env} and \S \ref{sec:env}), the accretion rates
explored include 27 discrete values between 
$\sim 10^{-7}-10^{-4}$ M$_{\odot}$ yr$^{-1}$. Sampled
values of the other parameters are $R_{\rm c}=30,100$, and $500$ AU; 
$R_{\rm out}=500,1000$, and $2000$ AU;
and $i=5-90^{\circ}$ in increments of $5^{\circ}$.
For the pure disk model, we vary the disk mass, scale height,
outer radius, and inclination; $M_{\rm disk}=10^{-3},
5 \times 10^{-2},$ and $1.0$ M$_{\odot}$, $h_0 = 5,15$, and $25$ AU,
$R_{\rm out}=500,1000$, and $2000$ AU,
and $i=5-90^{\circ}$ in increments of $5^{\circ}$ .  
Finally, for the envelope+disk
model, we assume $h_0=15$ AU, but allow $M_{\rm disk}$
to vary, in addition to varying several important envelope parameters:
$\dot{M}$, $R_{\rm c}$, $R_{\rm out}$, and $i$.  The range of envelope
parameters are the same as for the pure envelope model, and we sample
$M_{\rm disk} = 10^{-3}, 10^{-2}$, and 1.0 M$_{\odot}$.  All of the parameter
values in these grids are chosen to bracket physically-plausible values.

In order to compare with our observations, we use MC3D to compute 
self-consistently the temperature distribution, the SED from 0.1-5000 $\mu$m, 
the scattered light image at 
0.9 $\mu$m, and the dust continuum image at 1.3 mm, for each model.  
The best-fit models are determined by minimizing the residuals between model 
and data for the combined SED+imaging dataset. Specifically, we minimize the
quantity 
\begin{equation}
\chi_{\rm r,tot}^2 \equiv \chi_{\rm r, SED}^2 + \chi_{\rm r, 0.9 \mu m}^2
+ \chi_{\rm r,1 mm}^2.
\label{eq:x2_norm1}
\end{equation}
Here, $\chi_{\rm r, SED}^2$ is the reduced $\chi^2$ value of a model fitted 
only to the SED data, and $\chi_{\rm r, 0.9 \mu m}^2$ and $\chi_{\rm r,1 mm}^2$
are the analogous quantities for the $I$-band and 1 mm images.  Thus, 
$\chi^2_{\rm r,tot}$ gives equal weight to the three datasets (as opposed
to a true $\chi_{\rm r}^2$ for the combined SED+imaging dataset, which would
give more weight to the greater number of imaging datapoints relative
to SED measurements).

To calculate $\chi_{\rm r}^2$ values for each dataset, we estimate
the uncertainties by assuming that the best-fit model (amongst all
classes of models considered) for a given dataset has 
a reduced $\chi^2$ value of unity, and setting the error bars accordingly.  
This means that when a model is fitted to a given dataset (e.g., SEDs), 
the reduced $\chi^2$ is given by the sum of the
squared residuals divided by the minimum value of this quantity for all
of the models considered: denoting summed squared residuals by $X^2$, 
$\chi_{\rm r}^2 = X^2/{\rm min}(X^2)$.
Thus, Equation \ref{eq:x2_norm1} can be re-written as
\begin{equation}
\chi^2_{\rm r,tot} = \frac{X^2_{\rm sed}}{{\rm min}(X^2_{\rm sed})} + 
\frac{X^2_{\rm 0.9 \mu m}}{{\rm min}(X^2_{\rm 0.9 \mu m})} + 
\frac{X^2_{\rm 1mm}}{{\rm min}(X^2_{\rm 1mm})}.
\label{eq:x2_norm}
\end{equation}
As mentioned above, when fitting the SED we minimize the difference between
the logarithms of the model and data, to emphasize the shape of the SED
over the peak value.  In addition, we allow a scale factor between
the model SED and the observed fluxes.  When fitting 0.9 $\mu$m and 1 mm 
images, we first rotate
the observed images to a position angle where the brightest scattered light
is south of the millimeter emission, consistent with the position
angle definition in our models.

Once the best-fit
model out of these grids has been determined, we attempt to ``zoom in'' on
the solution with more finely gridded values of inclination.   During this
zoom-in stage, we also vary $L_{\rm central}$ to produce the correct
normalization between the modeled and observed SEDs (without a scale factor).
Because the central luminosity is somewhat degenerate with the total
optical depth of the system \citep[e.g.,][]{KCH93}, minor adjustments to
$\dot{M}$ and/or $M_{\rm disk}$ may be required when 
$L_{\rm central}$ is varied.
Thus, when zooming in on a solution, we vary $i$, $L_{\rm central}$, and
$\dot{M}$ and/or $M_{\rm disk}$ simultaneously.


The effects of various parameters on the SEDs have already been
explored by previous investigators 
\citep[e.g.,][]{ALS87,KCH93,DALESSIO+99}, and some work
has studied the effects on scattered light images 
\citep[e.g.,][]{WH92,WKG97}.  We attempt to
improve on this previous work by illustrating the effects of model parameters
on our combined SED+imaging dataset for a broad range of circumstellar 
geometries, and discussing how degeneracies between
fitted parameters can be broken.  In the remainder of this section, we
describe the rotating infalling envelope, flared disk, and disk+envelope
geometries.  In the following section, we present the best-fit model from
each category for comparison to our observations.

\subsection{Rotating, Infalling Envelope \label{sec:env}}
The density distribution for a rotating, infalling envelope is given by
\citep[e.g.,][]{ULRICH76,CM81,TSC84},
\begin{equation}
\rho_{\rm env} (r,\theta) = \frac{\dot{M}}{4\pi} (GM_{\ast}r^3)^{-1/2}
\left(1 + \frac{\mu}{\mu_0}\right)^{-1/2} \left(\frac{\mu}{\mu_0} + 2\mu_0^2
\frac{R_c}{r}\right)^{-1}.
\label{eq:env}
\end{equation}
Here, $r$ is the radial coordinate,
$M_{\ast}$ is the mass of the central star (assumed to be 0.5 M$_{\odot}$), 
$\dot{M}$ is the
mass infall rate, $R_c$ is the centrifugal radius, $\mu=\cos\theta$ 
defines the angle above the midplane,
and $\mu_0$ defines the initial streamline of the infalling material.

For each value of $r,\theta$, there is a unique value of $\mu_0$, since
the streamlines of infalling particles do not cross.
Before determining the density distribution using Equation \ref{eq:env},
we analytically solve for $\mu_0$ :
\begin{equation}
\frac{r}{R_c} = \frac{\cos \theta_0 \sin^2 \theta_0}{\cos \theta_0 - 
\cos \theta} = \frac{1-\mu_0^2}{1-\mu/\mu_0}.
\end{equation}
This is a cubic equation with three solutions.  The correct solution is the
one for which $\sin \theta_0$ has the same sign as $\sin \theta$, and for
which the solution is real.  In other words, a particle that
starts in the northern or southern hemisphere never leaves that hemisphere.
The analytic solution is thus
\begin{equation}
\mu_0 = \xi - \frac{1}{3\xi},
\end{equation}
where $\xi$ is defined as
\begin{equation}
\xi \equiv \frac{\left(27\mu r R_c^2 + \sqrt{729 \mu^2 r^2 R_c^4 + 108 R_c^3
(r-R_c)^3}\right)^{(1/3)}}{2^{(1/3)} 3 R_c}.
\end{equation}

We include an outflow cavity in our envelope models by decreasing the
density by some factor, $f_{\rm cav}$, 
in the polar regions of the circumstellar distributions.
The cavity shape is defined by
\begin{equation}
\rho (r,z>z_0 + r^\zeta) =  f_{\rm cav} \times \rho_{\rm env}(r,z).
\end{equation}
Here, $z$ is the height above the midplane,
$z_0$ describes how close to the star the outflow cavity begins,
and $\zeta$ describes the opening angle and shape of the outflow.
We fix $z_0 = 1$ AU, since this parameter does not have major affects
on the observed SEDs or scattered light images.  We also assume that 
within the outflow cavity, the density decreases by a factor of $f_{\rm cav}=4$
relative to the rest of the envelope; this value of $f_{\rm cav}$ produces the
correct amount of extinction to match the observed short-wavelength SEDs.
The cavity profile has some effect on the structure of modeled scattered light 
emission. However, these effects are
small compared with the other envelope properties; thus, we fix the
value of $\zeta$ at 1.1 in our analysis.  Using other types of cavities (e.g.,
cones) may have small effects on the scattered light images, but will
have negligible effects on SEDs \citep[e.g.,][]{WHITNEY+03a}.

We fit this envelope model to our data by varying
$\dot{M}$, $R_{\rm c}$,
the outer radius of the envelope, $R_{\rm out}$, and the 
viewing angle of the model, $i$.  Some of these parameters produce
degenerate effects on the models, which is why we generate grids
where these parameters are varied simultaneously.  However, we address
the effects of each parameter below in order to shed light on how
our different datasets constrain various envelope properties.

\begin{figure}
\plotone{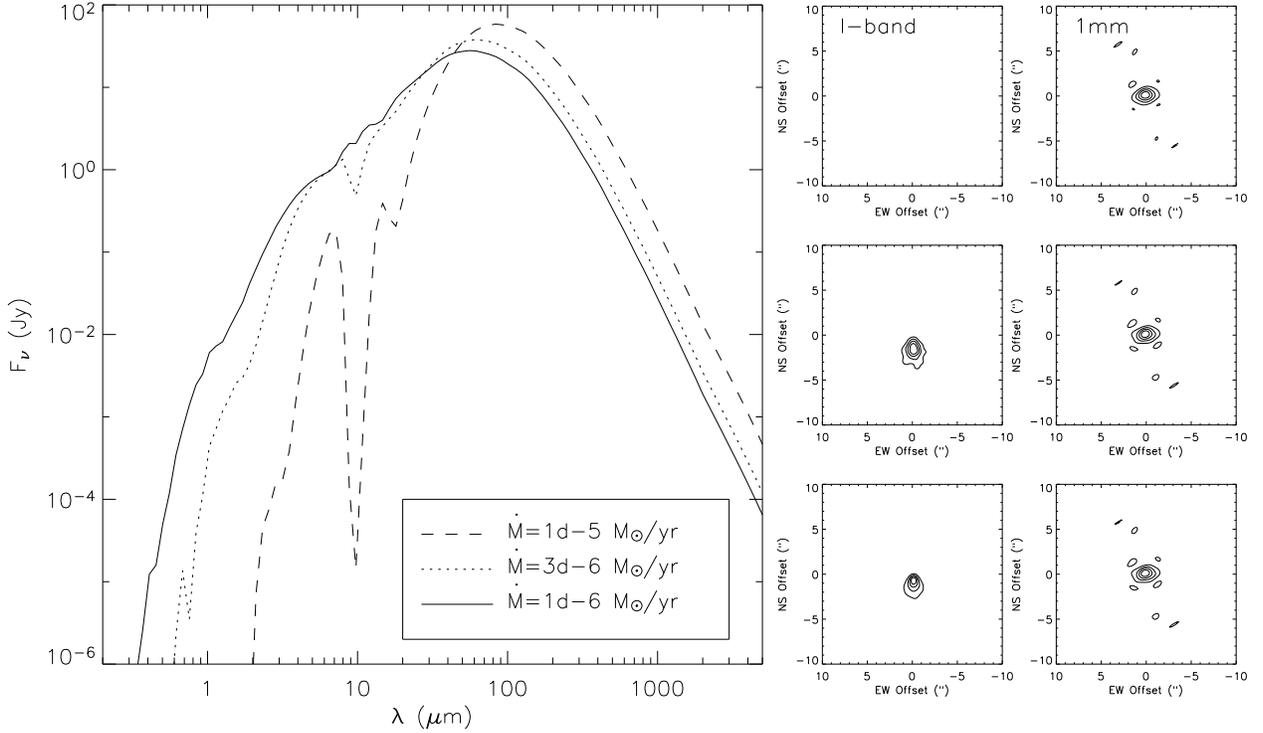}
\caption[Dependence of envelope model on $\dot{M}$]
{SEDs, scattered light images, and millimeter continuum images
for a rotating, collapsing envelope with a range of
mass accretion rates, spanning $10^{-6}$ to
$10^{-5}$ M$_{\odot}$ yr$^{-1}$
($\dot{M}$ increases from the bottom to top panels).  This model assumes
$R_{\rm c}=30$ AU, $R_{\rm out}=1000$ AU, and $i=45^{\circ}$.
The $I$-band images were generated by convolving our model images
with a $1''$ FWHM Gaussian, to simulate the seeing of our LRIS observations.
Similarly, we convolved the 1 mm images with the actual beam that was used to
observe IRAS 04016+2610. For both 0.9 $\mu$m and 1 mm images, the contour
intervals are 20\% of the peak flux, beginning at 20\%.  
The noisy appearance of some of the model images in this and subsequent
figures is due to the finite number of photons used in our Monte Carlo
radiative transfer modeling, and does not reflect underlying substructure.
As $\dot{M}$ increases,
the strength of near-IR scattered emission decreases, although the shape
of the emission broadens (note that for $\dot{M}=10^{-5}$ M$_{\odot}$ 
yr$^{-1}$, the scattered emission is undetected in our models).  
Higher mass accretion rates also produce larger
mid-IR absorptions.  While the appearance of millimeter images is
relatively unaffected by changes in $\dot{M}$, the long-wavelength
flux is enhanced for higher accretion rates.
\label{fig:env_mdot}}
\end{figure}

The accretion rate, $\dot{M}$, is
one of the most important parameters in terms of its effects on the
SED, since it is linked to the total envelope mass (Equation \ref{eq:env}),
and thus governs directly the optical depth of the dust
surrounding the protostar (Figure \ref{fig:env_mdot}).  
For higher mass accretion rates, the
optical depth will increase, and less of the short-wavelength emission
will escape (either directly or as scattered light).  
Moreover, the mid-IR absorption deepens, because of increased extinction of
the stellar emission and absorption in the 10 and 18 $\mu$m silicate features.
The short-wavelength radiation is re-processed and emitted at longer
wavelengths, leading to more emission at far-IR through millimeter 
wavelengths for higher values of $\dot{M}$.  The higher optical depths
associated with large mass accretion rates also push the visible scattering
surfaces outward, leading to somewhat larger $I$-band images.  However,
for very large accretion rates, the near-IR emission may be quenched 
altogether (Figure \ref{fig:env_mdot}).

\begin{figure}
\plotone{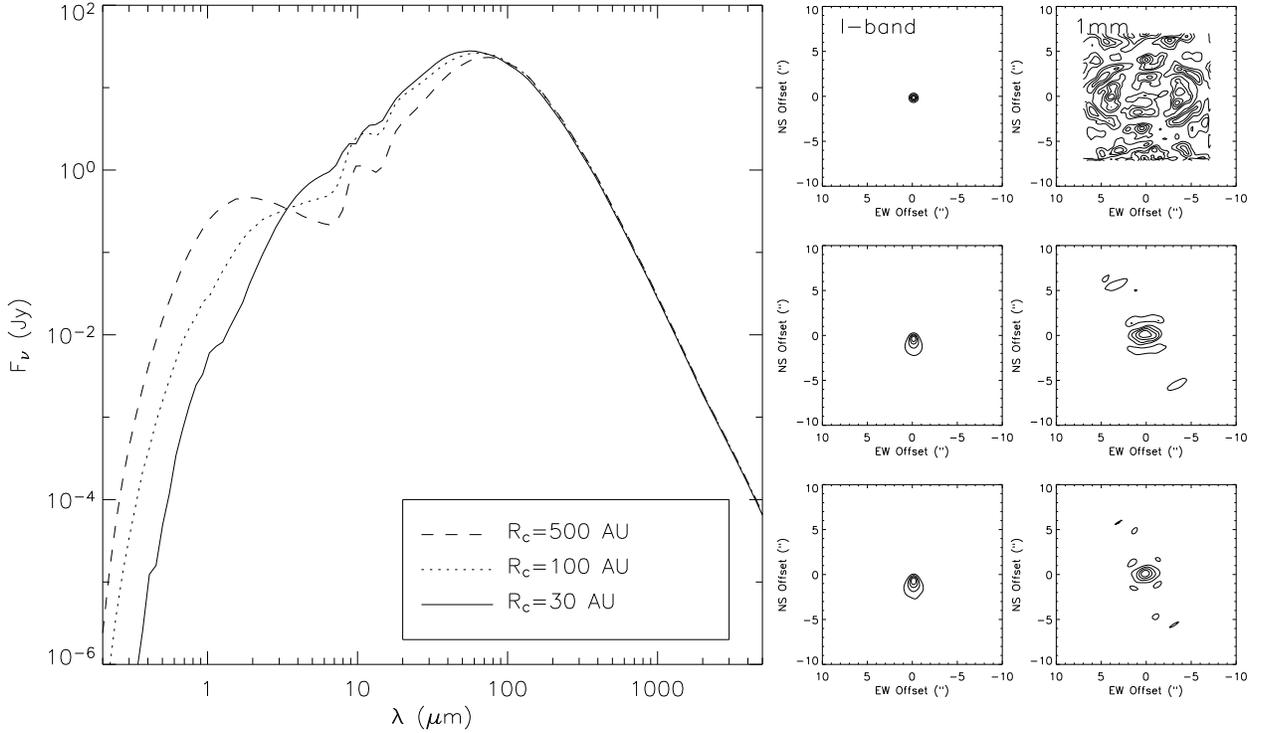}
\caption[Dependence of envelope model on $R_{\rm c}$]
{SEDs, scattered light images, and millimeter continuum images
for a rotating, collapsing envelope with a range of
centrifugal radii, spanning 30 to 500 AU 
($R_{\rm c}$ increases from the bottom to top panels).
This model assumes $\dot{M}=10^{-6}$ M$_{\odot}$ yr$^{-1}$, 
$R_{\rm out}=1000$ AU, and $i=45^{\circ}$.  Larger values of $R_{\rm c}$ lead
to less spherically symmetric dust distributions, leading to increased
absorption in the midplane, but less absorption out of the midplane.  Thus,
for large $R_{\rm c}$, the central star becomes visible directly.
The centrifugal radius also represents a region of enhanced density, and
thus larger $R_{\rm c}$ lead to significantly larger millimeter 
emission regions.
In fact for $R_{\rm c}=500$ AU, the millimeter emission is almost completely
over-resolved in our OVRO observations.
\label{fig:env_rc}}
\end{figure}

The centrifugal radius, $R_{\rm c}$, defines the radius in the model where
material falling in from outer regions joins rotationally-supported material
in the inner region.  Outside of $R_{\rm c}$, the envelope is essentially
spherical (except for the outflow cavity), while the density distribution
is significantly flattened interior to the centrifugal radius.  In addition,
matter builds up at $R_{\rm c}$, leading to a region of enhanced density.  
For smaller values of $R_{\rm c}$, 
a larger fraction of the envelope material is spherically distributed,
and the optical depth is larger for most sight-lines.  Thus, models with
small $R_{\rm c}$ show large near-IR extinctions and
deep mid-IR absorption
(largely degenerate with the effects of $\dot{M}$; Figures \ref{fig:env_mdot}
and \ref{fig:env_rc}).
In contrast,
for larger $R_{\rm c}$, more of the envelope's mass is relegated to the
midplane, making it easier for emission to escape.  Thus, scattered
emission is stronger for larger values of $R_{\rm c}$.  This parameter
also has a critical effect on the modeled millimeter images: since 
the density is higher for radii $\la R_{\rm c}$, the size of the
millimeter emission will correlate with the centrifugal radius 
(Figure \ref{fig:env_rc}).  Thus,
the sizes of our millimeter images constrain $R_{\rm c}$ directly.

\begin{figure}
\plotone{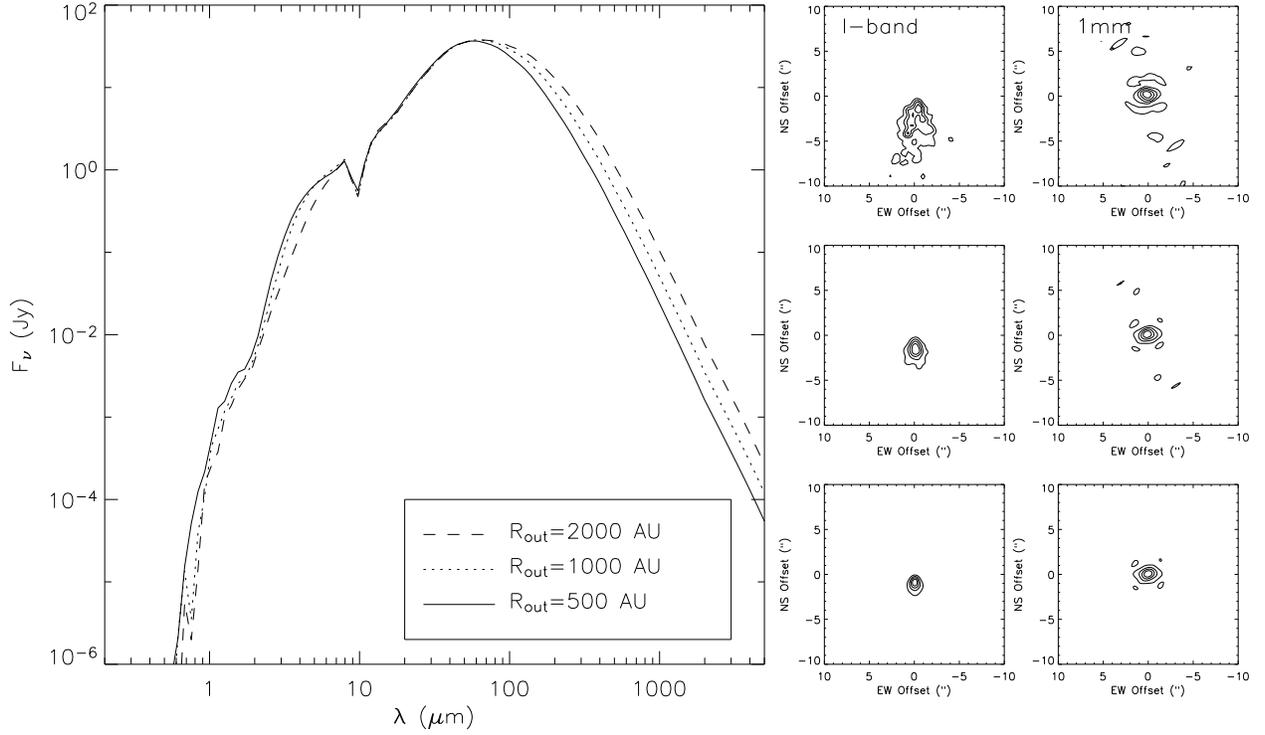}
\caption[Dependence of envelope model on $R_{\rm out}$]
{SEDs, scattered light images, and millimeter continuum images
for a rotating, collapsing envelope with a range of outer radii, spanning 
500 to 2000 AU ($R_{\rm out}$ increases from the bottom to top panels).  
This model assumes
$\dot{M}=3 \times 10^{-6}$ M$_{\odot}$ yr$^{-1}$, 
$R_{\rm c}=30$ AU, and $i=45^{\circ}$.
As $R_{\rm out}$ increases, the amount of material in the envelope is increased
(for a fixed $\dot{M}$), increasing the optical depth and leading to 
slightly higher
absorption at near-to-mid-IR wavelengths.  Not surprisingly, the images
at both 0.9 $\mu$m and 1 mm become larger as $R_{\rm out}$ is increased.
For the 1 mm images, the increased extended emission is manifested primarily
by stronger sidelobe features, which are artifacts of the limited sampling
in our interferometric observations (the 1 mm images have been convolved with
the OVRO beam obtained for IRAS 04016+2610). 
\label{fig:env_rout}}
\end{figure}

The outer radius also affects the model, mainly by changing the
optical depth; larger values of $R_{\rm out}$ lead to more mass in the
envelope, and thus more extinction of short wavelength light and
more emission of long-wavelength radiation (Figure \ref{fig:env_rout}).  
Smaller values of
$R_{\rm out}$ may require larger values of $\dot{M}$ and/or smaller
values of $R_{\rm c}$ in order to correctly model the SED.
$R_{\rm out}$ is constrained strongly from our scattered light and millimeter
images, since larger outer radii produce larger images (Figure 
\ref{fig:env_rout}).  For our 
sample, the observed sizes of scattered light constrain $R_{\rm out}$ to be
$\ga 1000$ AU, while the fact that the millimeter images are fairly compact
rules out outer radii larger than $\sim 1500-2000$ AU.

\begin{figure}
\plotone{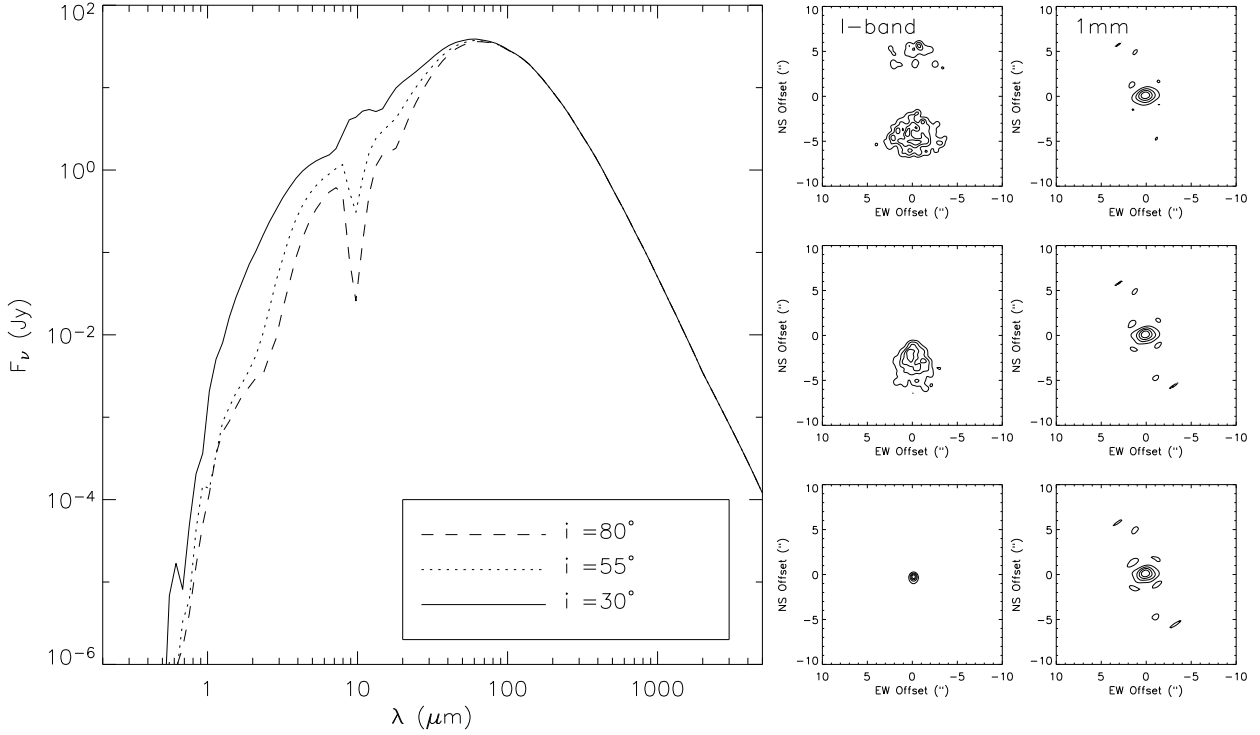}
\caption[Dependence of envelope model on inclination]
{SEDs, scattered light images, and millimeter continuum images 
for a flared disk model at a range of viewing angles
($i$ increases from the bottom to top panels).  
More edge-on models exhibit deeper absorption at mid-IR wavelengths, and
higher extinction of the central star.  For small inclinations
($i \la 30^{\circ}$), the central star is visible, and dominates the
$I$-band emission.  For moderate inclinations, an asymmetric scattered light
structure is observed, while for nearly edge-on orientations, a symmetric,
double-lobed structure is observed.  The millimeter emission for this model
is unresolved by our observations, and the model images are thus insensitive
to inclination.
\label{fig:env_inc}}
\end{figure}

Inclination is a crucial parameter for modeling both the SED and
the scattered light images.  
As noted by previous authors \citep[e.g.,][]{KCH93,NNU03} and shown in Figure 
\ref{fig:env_inc},
the SED changes drastically depending on viewing angle. At larger
inclinations, where the observer's line of sight passes through more
of the dense midplane, the overall optical depth of the model increases,
leading to enhanced extinction at short wavelengths and absorption at
mid-IR wavelengths.  In contrast,
for smaller inclinations the optical depth is lower, and for face-on
models, the line of sight may go directly down the outflow cavity to
the central protostar.  In addition to affecting the amount of scattering
or direct stellar radiation, the inclination also affects the shape
of the scattered emission: for edge-on distributions, the scattered light
arises mainly from the edges of the outflow cavity, producing a symmetric
structure with two lobes.  As the
inclination decreases, one lobe brightens with respect to the other, and
moves closer to the central protostar 
as the line of sight pierces further into the
cavity.  Finally, for inclinations close to face-on, the protostar is
visible directly, swamping any scattered emission that might be present.
Since our millimeter images are only marginally spatially resolved, inclination
is not important in modeling this emission.  However, the offset between 
1 mm and 0.9 $\mu$m emission {\it is} sensitive to inclination; the
visible scattering surface moves farther from the mass-sensitive
millimeter emission for larger inclinations (Figure \ref{fig:env_inc}).

\subsection{Flared Disk \label{sec:disk}}
The density distribution for a flared disk is given by \citep{SS73},
\begin{equation}
\rho_{\rm disk} (r,z) = \rho_0 \left(\frac{R_{\ast}}{r}\right)^{\alpha} 
\exp\left\{-\frac{1}{2}\left[\frac{z}{h(r)}\right]^2\right\},
\end{equation}
where
\begin{equation}
h(r) = h_0 \left(\frac{r}{\rm 100 \: AU}\right)^{\beta}.
\end{equation}
Here, $r$ is the radial distance from the star in the disk mid-plane, 
$z$ is the vertical distance from the mid-plane, $R_{\ast}$ is the
stellar radius, and $h$ is the disk scale height.  We assume fixed
values for the exponents of radial and vertical density distributions,
$\alpha$ and $\beta$, respectively.  For a flared disk in hydrostatic
equilibrium, one expects $\beta=58/45$ \citep{CG97}, and we adopt this
value.  Using the relation that results from viscous accretion theory,
$\alpha=3 (\beta-1/2)$ \citep{SS73}, we obtain $\alpha=2.37$.  These
values for $\alpha$ and $\beta$ are similar to those used in previous
modeling of circumstellar disks \citep[e.g.,][]{DALESSIO+99,WOOD+02,WPS03}.
We fit the disk model to our data by varying M$_{\rm disk}$, $h_0$, 
$R_{\rm out}$, and the viewing angle of the model, $i$. 

\begin{figure}
\plotone{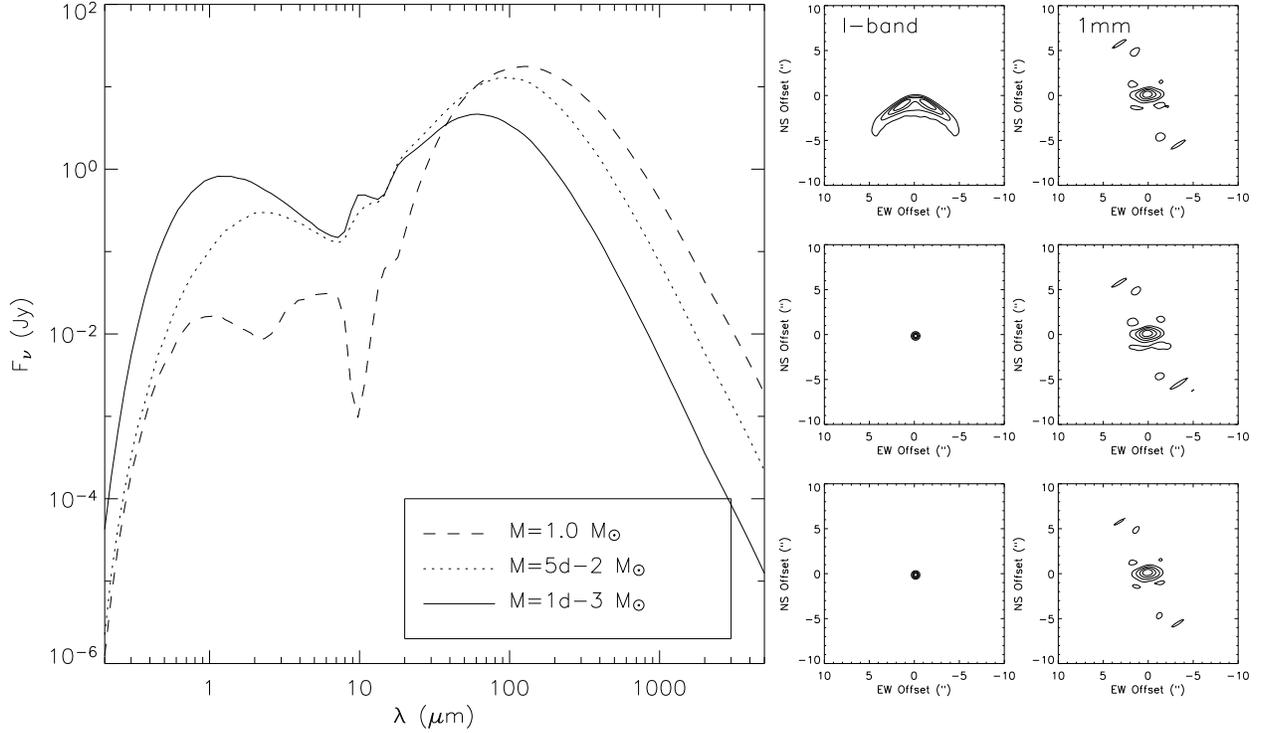}
\caption[Dependence of disk model on $M_{\rm disk}$]
{SEDs, scattered light images, and millimeter continuum images 
for a flared disk model with a range of disk masses
($M_{\rm disk}$ increases from the bottom to top panels).
For this model, $R_{\rm out}=1000$ AU, $h_0=15$ AU, and $i=60^{\circ}$.
Higher disk mass increases the optical depth of the model, leading to
higher absorption at short wavelengths and enhanced emission at
long wavelengths.  In addition, if there is sufficient dust mass,
the central star is obscured, and scattered light is visible at short 
wavelengths.  The flux in millimeter continuum images is increased for 
higher disk mass, and the emission becomes slightly more extended.
\label{fig:disk_mass}}
\end{figure}

The disk mass affects our modeled SEDs, scattered light images, and
millimeter images.  Regarding the SEDs, higher dust mass increases the 
optical depth, leading to higher obscuration of the central star, deeper 10 
and 18 $\mu$m silicate absorption, and enhanced re-emission at long 
wavelengths. As illustrated by Figure \ref{fig:disk_mass}, the disk mass also 
has substantial
effects on the scattered light: for sufficiently small disk masses 
($\la 10^{-2}$ M$_{\odot}$) the central protostar is
visible for inclination $\la 70^{\circ}$ and this stellar radiation
swamps any scattered light.  However, for higher disk masses, the flared
disk surface becomes sufficiently optically-thick to obscure the protostar,
allowing scattered emission to be observed.  The 
morphologies of millimeter images are
only slightly affected by disk mass: more massive disks have
millimeter emission detectable out to larger radii, leading to slightly 
larger observed images.

\begin{figure}
\plotone{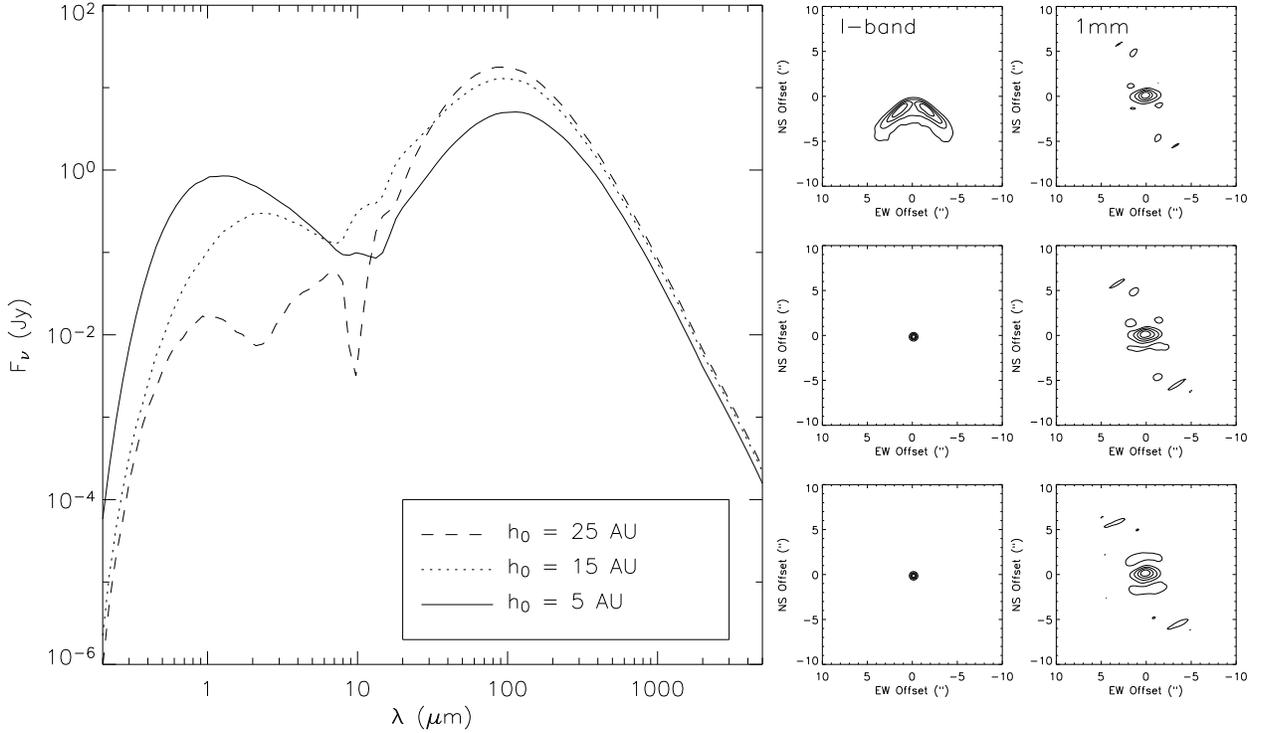}
\caption[Dependence of disk model on $h_0$]
{SEDs, scattered light images, and millimeter continuum images 
for a flared disk model with a range of disk scale heights
($h_{0}$ increases from the bottom to top panels).
This model assumes $M_{\rm disk}=5 \times 10^{-2}$ M$_{\odot}$,
$R_{\rm out}=1000$ AU, and $i=60^{\circ}$. Larger values of $h_0$ produce
similar effects on the short-wavelength SED and scattered light images
as larger values of $M_{\rm disk}$ (Figure \ref{fig:disk_mass}); higher
absorption, and increased visibility of scattered light.  However, the
effects on the longer-wavelength emission differ; higher values of $h_0$
increase the far-IR flux, but have little effect on millimeter fluxes 
and lead to slightly more compact morphologies at millimeter wavelengths
(corresponding to weaker sidelobe emission features).
\label{fig:disk_h0}}
\end{figure}

The effects of $h_0$ are similar to the effects of $M_{\rm disk}$ in 
some respects; however, SED+imaging data allows us to disentangle the two.
Larger values of $h_0$ produce more absorption at short wavelengths, 
and if the scale height becomes large enough to obscure the central protostar, 
scattered emission becomes visible (Figure \ref{fig:disk_h0}).  These effects 
are similar to those produced by higher values of disk mass.  However, the
effects on the longer-wavelength emission differ; higher values of $h_0$
increase the far-IR flux, but have little effect on fluxes or image
morphologies at millimeter wavelengths.

\begin{figure}
\plotone{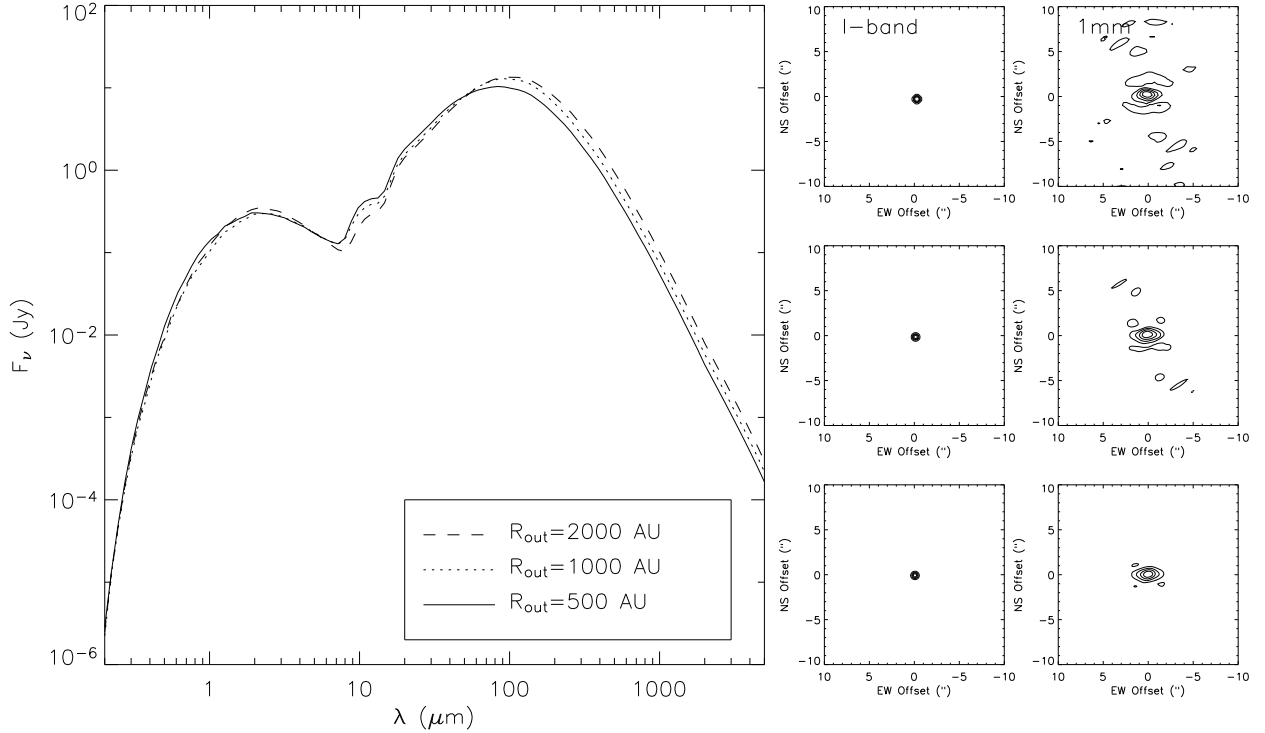}
\caption[Dependence of disk model on $R_{\rm out}$]
{SEDs, scattered light images, and millimeter continuum images 
for a flared disk model with a range of outer radii
($R_{\rm out}$ increases from the bottom to top panels).
This model assumes $M_{\rm disk}=5 \times 10^{-2}$ M$_{\odot}$,
$h_0=15$ AU, and $i=60^{\circ}$.  Since we have fixed the disk mass,
different values of $R_{\rm out}$ do not significantly affect the opacity
of the model.  However, a larger outer radius produces a larger surface
area of cool dust, which leads to a slight enhancement of the long-wavelength
flux.  Moreover, larger $R_{\rm out}$ will produce larger images
in both scattered light and millimeter emission (the short-wavelength
images shown here are dominated by light from the central protostar,
and thus no scattered emission is visible).
\label{fig:disk_rout}}
\end{figure}

Changing $R_{\rm out}$ does not significantly affect the SEDs, as long as
the disk mass is held fixed (Figure \ref{fig:disk_rout}).  This is because
the optical depth of the model is not affected substantially by $R_{\rm out}$.
However, there are small changes due to the larger surface area of
optically-thick dust for larger values of $R_{\rm out}$.  In addition,
larger outer radii produce larger images in both scattered light and 
millimeter emission, since disk material is distributed to larger radii.

Disk viewing angle is a crucial property in modeling the SEDs and scattered
light images \citep[e.g.,][]{DALESSIO+99,WH92}.  
Similar to the dependence of envelope model SEDs on inclination, 
more edge-on disk models exhibit deeper absorption at mid-IR wavelengths, and
higher extinction of the central star (Figure \ref{fig:disk_inc}).  
Inclination also substantially affects the observed scattered light:
for moderate inclinations ($i \la 55^{\circ}$), the central star is visible 
and dominates the $I$-band emission, while for larger inclinations the 
protostar is obscured and an asymmetric scattered light structure is observed. 
For nearly edge-on orientations, a symmetric, structure is observed, 
corresponding to the top and bottom surfaces of the flared disk. As for
envelope models, our marginally resolved millimeter images can not constrain
well the inclination, although the offset between scattered light and
millimeter emission does probe the orientation of disk models.  

\subsection{Envelope+Disk \label{sec:d+e}}
\begin{figure}
\plotone{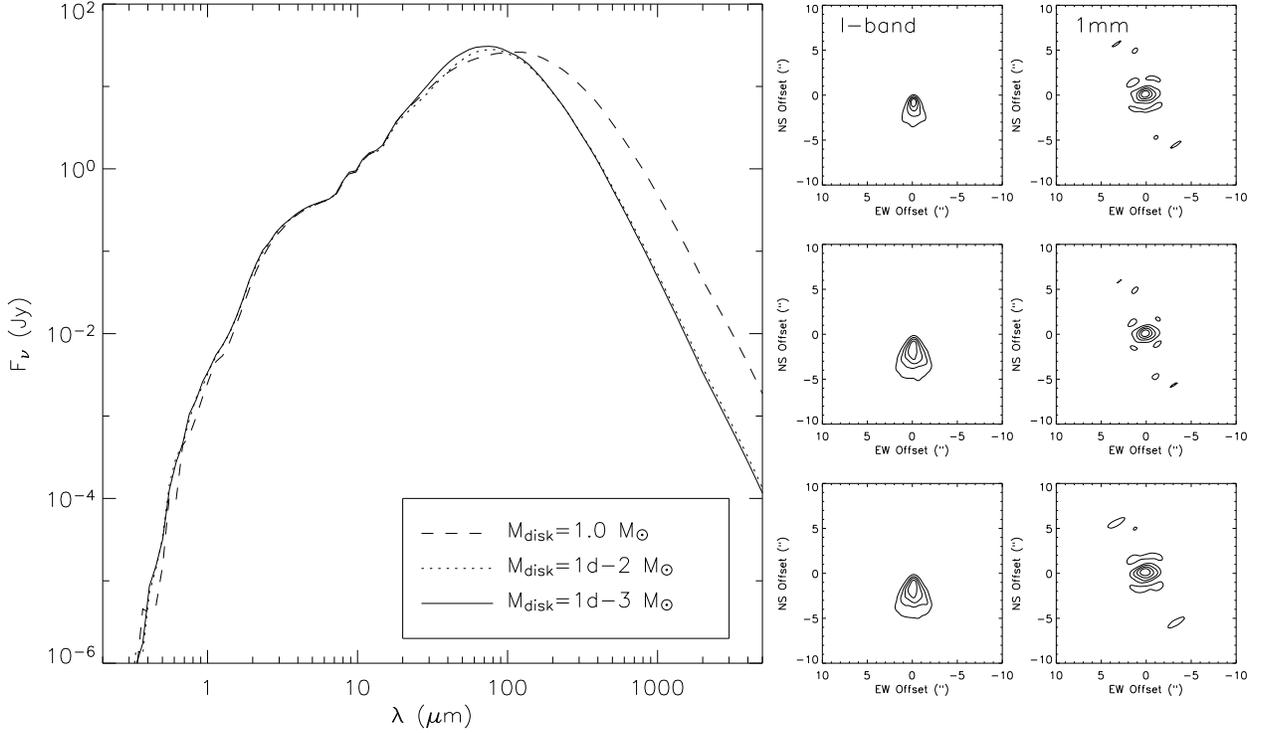}
\caption[Dependence of disk+envelope model on $M_{\rm disk}$]
{SEDs for envelope+disk models for a range of disk masses
($M_{\rm disk}$ increases from the bottom to top panels).
The inclination of the models has been fixed at 45$^{\circ}$.
The envelope component of the model has the following properties:
$\dot{M}=3 \times 10^{-6}$ M$_{\odot}$ yr$^{-1}$,
$M_{\ast} = 0.5$ M$_{\odot}$, $R_{\rm c}=100$ AU, and $R_{\rm out}=1000$ AU.
The various models illustrate the effect of adding progressively more
massive (plotted bottom to top) disks to the model: 
$M_{\rm disk}=10^{-3}$, $10^{-2}$, and $1.0$ M$_{\odot}$.  The main effect
of the disk is on the SED long-ward of 100 $\mu$m, where the dense, cool
dust in the disk component adds substantial flux.  The 0.9 $\mu$m and 1 mm
images are also affected slightly, becoming more centrally concentrated
for larger disk masses.
\label{fig:sed_disk_env}}
\end{figure}

In addition to pure envelopes and pure disks, we consider a model
incorporating both an envelope and an embedded disk.  
The main difference between this model and the pure envelope model is an 
enhanced density in the midplane, especially interior to the centrifugal 
radius. As a result of this enhanced midplane density, more long-wavelength
emission is produced than for a pure envelope model.  Since the amount
of long-wavelength emission for a pure envelope model depends primarily
on the mass infall rate, $\dot{M}$, additional long-wavelength emission
from the disk component may allow good fits to the data with lower
inferred values of $\dot{M}$.  The effect of the disk component on the
SED is illustrated in Figure \ref{fig:sed_disk_env}.  

In addition to larger fluxes, inclusion of a disk component also leads to
more centrally concentrated modeled scattered light and millimeter images.
With our marginally resolved
millimeter images, we can not probe such density concentrations directly.
However, as we discuss further below, comparison of our interferometrically
measured fluxes, which trace compact emission, with lower-resolution 
measurements \citep[e.g.,][]{MA01,YOUNG+03} can provide some constraint on the
relative amounts of compact and large-scale material, providing an additional
test of how well these models fit the data.

The envelope+disk model is implemented by computing the density distributions 
for both a disk and an envelope, and then setting the density of the combined
model to be the greater of the two individual densities for
a given position, $(r, \theta)$.  The temperature
distribution, spectral energy distribution, and images are then calculated
self-consistently for the combined model using MC3D.  In our current
implementation of the disk+envelope density distribution, 
the outer radius of the disk cannot be specified independently
of the outer radius of the envelope.  Thus, our models may not truly 
represent a physical disk+envelope model, where one might expect the disk
component to end at $R_c$.  These models should therefore be regarded
as only qualitative indicators that both disk and envelope components are 
needed to match the data.

\subsection{Disk+Extinction Model \label{sec:d+ex}}
Finally, we consider another variation of a disk+envelope model,  
incorporating a flared disk density distribution plus foreground extinction.  
The extinction in this model mimics the obscuring nature of the envelope
in the disk+envelope model described in \S \ref{sec:d+e}, 
but employs a simpler geometry.
In the following sections, we refer to this variation as a 
disk+extinction model.  

We consider values of $A_{\rm V}$ ranging from
0 to 60 mag, which allows foreground extinctions much higher than expected
from ambient material in the Taurus region \citep[e.g.,][]{KH95}.  
Extinctions higher than possible from ambient cloud material
($A_{\rm V} \sim 5$) probably arise in  material associated with the 
collapsing cloud cores surrounding the sources.  
Sub-millimeter fluxes observed in large-beam SCUBA 
maps \citep{YOUNG+03} can be used to estimate column densities, and thus 
extinctions toward our sample of Class I objects.  Since higher extinctions
are inferred for more compact dust distributions, models requiring larger
values of $A_{\rm V}$ imply smaller dust structures.  Based on the SCUBA
fluxes \citep{YOUNG+03},
extinctions larger than $\sim 5$ mag are only possible if the material is
distributed primarily on scales smaller than $\sim 5000$ AU, on the order
of expected envelope sizes.  

In this disk+extinction model, the morphologies of scattered light 
and millimeter continuum images are unaffected by the uniformly-distributed
obscuring envelope material, and thus the model images resemble those
predicted by pure disk models.  Since the dependence of the modeled images
on various disk parameters is the same as for pure disk
models (\S \ref{sec:disk}), we refer to Figures 
\ref{fig:disk_mass}--\ref{fig:disk_inc}.
However, the foreground extinction substantially alters 
the synthetic SEDs by eliminating much of the
short-wavelength stellar flux observed for pure disks.  
This model thus provides a way to produce disk-like images at the same time as 
heavily reddened SEDs.

\section{RESULTS \label{sec:results}}
\begin{figure}
\plotone{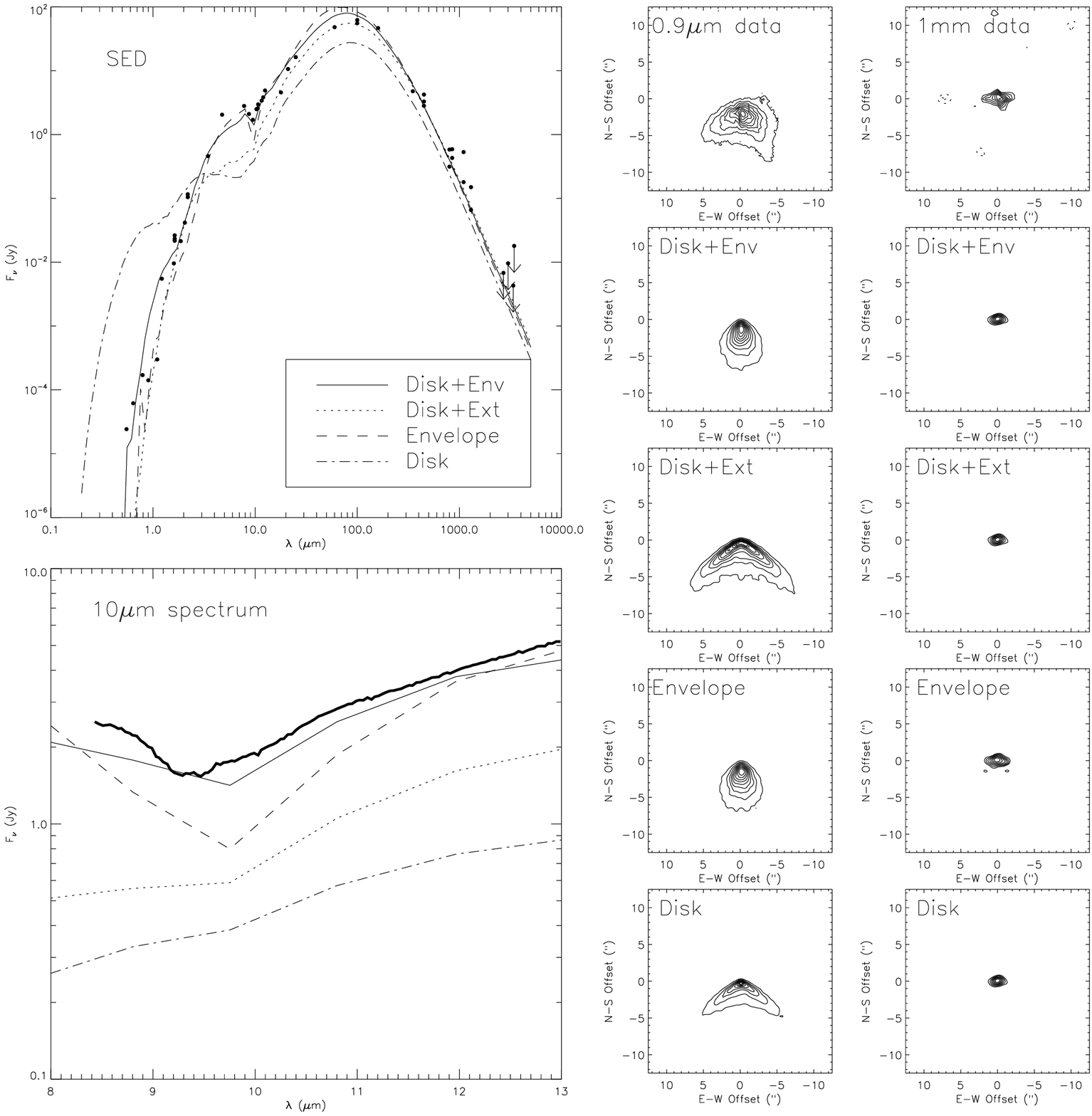}
\caption[Best-fit models for IRAS 04016+2610]
{SED data (Table \ref{tab:seds}), 10 $\mu$m spectrum 
\citep{KESSLER-SILACCI+05}, 0.9 $\mu$m scattered light image, 
and 1.3 mm continuum image for IRAS 04016+2610, along with best-fit models for
different assumed circumstellar geometries (Table \ref{tab:bestfits}).  
The observed images have been rotated on the sky by $-10^{\circ}$ 
(east of north) so that they have the same position angle definition as the 
models.  Images have contour levels of 10\% of the peak flux,
beginning at 10\% ($\sim 6\sigma$) for the 0.9 $\mu$m images and 
30\% ($\sim 2.5 \sigma$) for the 1 mm images. 
While pure disk models do not provide an adequate match to the
data, extincted disks, pure envelopes, and disk+envelopes match reasonably
well.  The disk+envelope, which provides the smallest residuals
between model and data, is preferred.
\label{fig:i04016}}
\end{figure}

\begin{figure}
\plotone{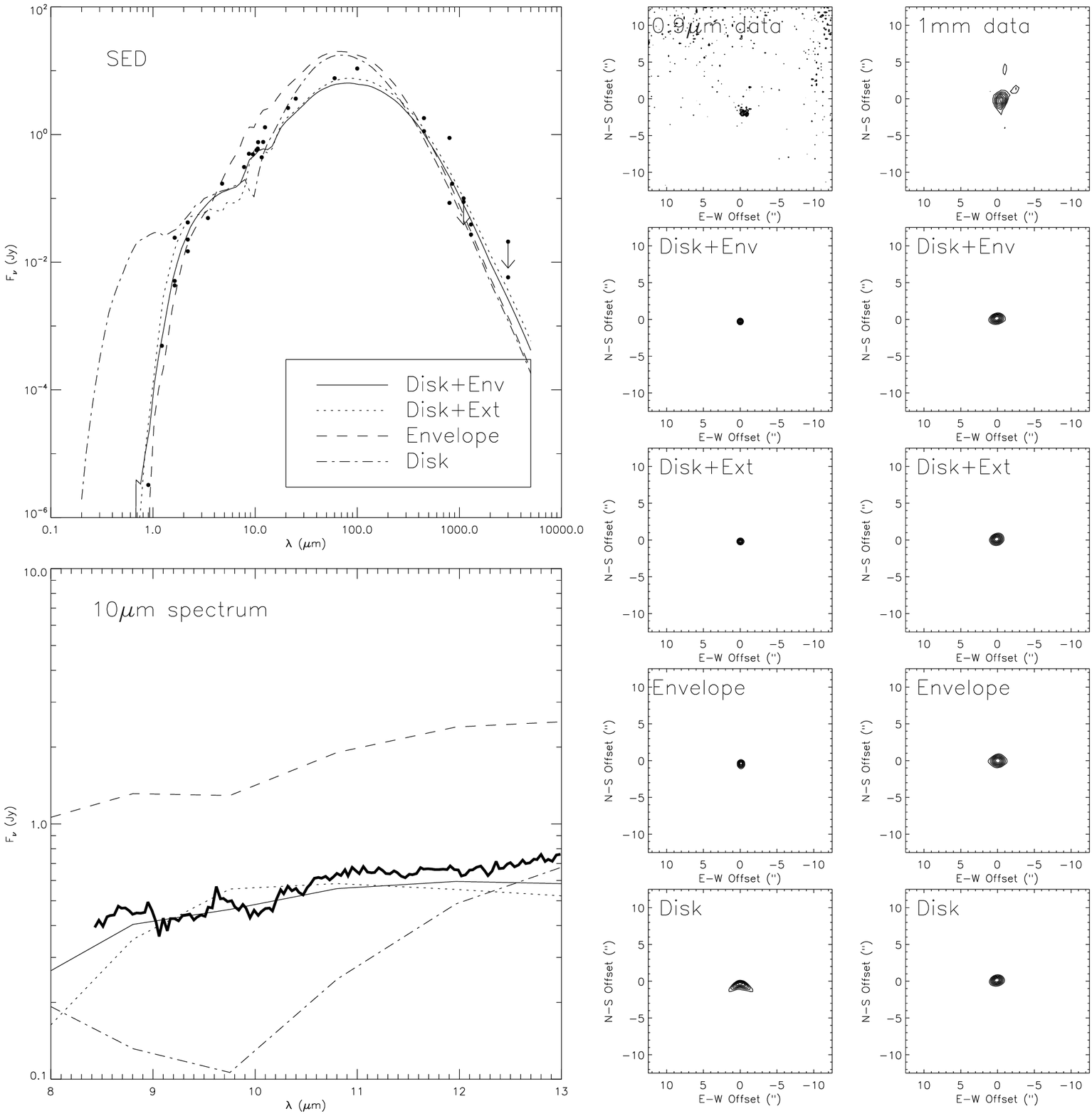}
\caption[Best-fit models for IRAS 04108+2803B]
{SED data (Table \ref{tab:seds}), 10 $\mu$m spectrum 
\citep{KESSLER-SILACCI+05}, 0.9 $\mu$m scattered light image, 
and 1.3 mm continuum image for IRAS 04108+2803B, along with best-fit models for
different assumed circumstellar geometries (Table \ref{tab:bestfits}).  
The observed images have been rotated on the sky by $-90^{\circ}$ 
(east of north) so that they have the same position angle definition as the 
models.  Images have contour levels of 10\% of the peak flux,
beginning at 30\% ($\sim 2.5 \sigma$) 
for the 0.9 $\mu$m images and 30\% ($\sim 3 \sigma$) for the 1 mm images.
Disk+envelope and disk+extinction models provide the best match to the data
for this source.
\label{fig:i04108}}
\end{figure}

\begin{figure}
\plotone{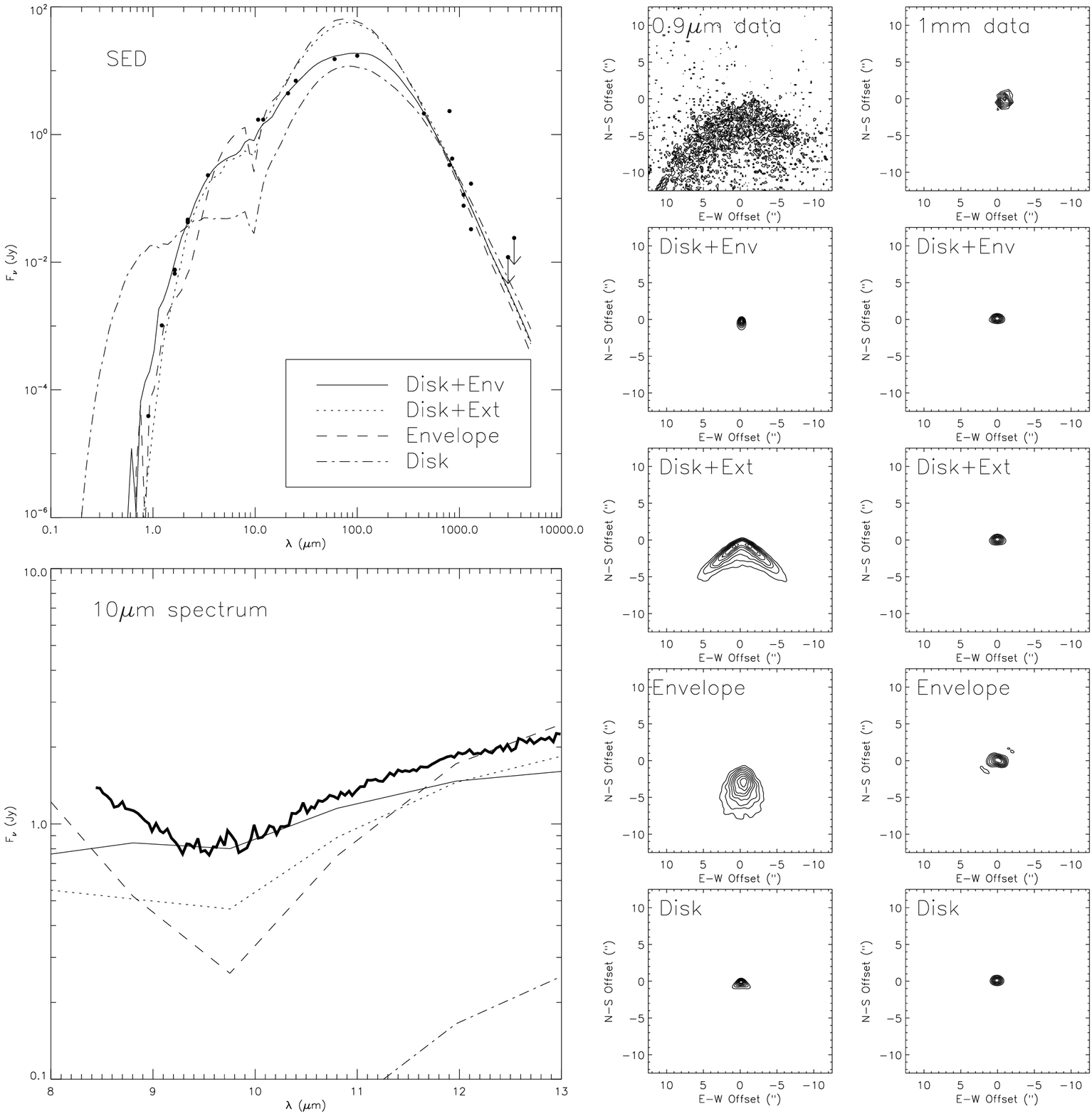}
\caption[Best-fit models for IRAS 04239+2436]
{SED data (Table \ref{tab:seds}), 10 $\mu$m spectrum 
\citep{KESSLER-SILACCI+05}, 0.9 $\mu$m scattered light image, 
and 1.3 mm continuum image for IRAS 04239+2436, along with best-fit models for
different assumed circumstellar geometries (Table \ref{tab:bestfits}).  
The observed images have been rotated on the sky by $-120^{\circ}$ 
(east of north) so that they have the same position angle definition as the 
models.  Images have contour levels of 10\% of the peak flux,
beginning at 20\% ($\sim 2.5 \sigma$) 
for the 0.9 $\mu$m images and 30\% ($\sim 5 \sigma$) for the 1 mm images.
For this object, the best-fit to the complete data set is obtained
for disk+extinction and disk+envelope models, although the SED and 10 $\mu$m 
spectra are fitted best by the disk+envelope model.
\label{fig:i04239}}
\end{figure}

\begin{figure}
\plotone{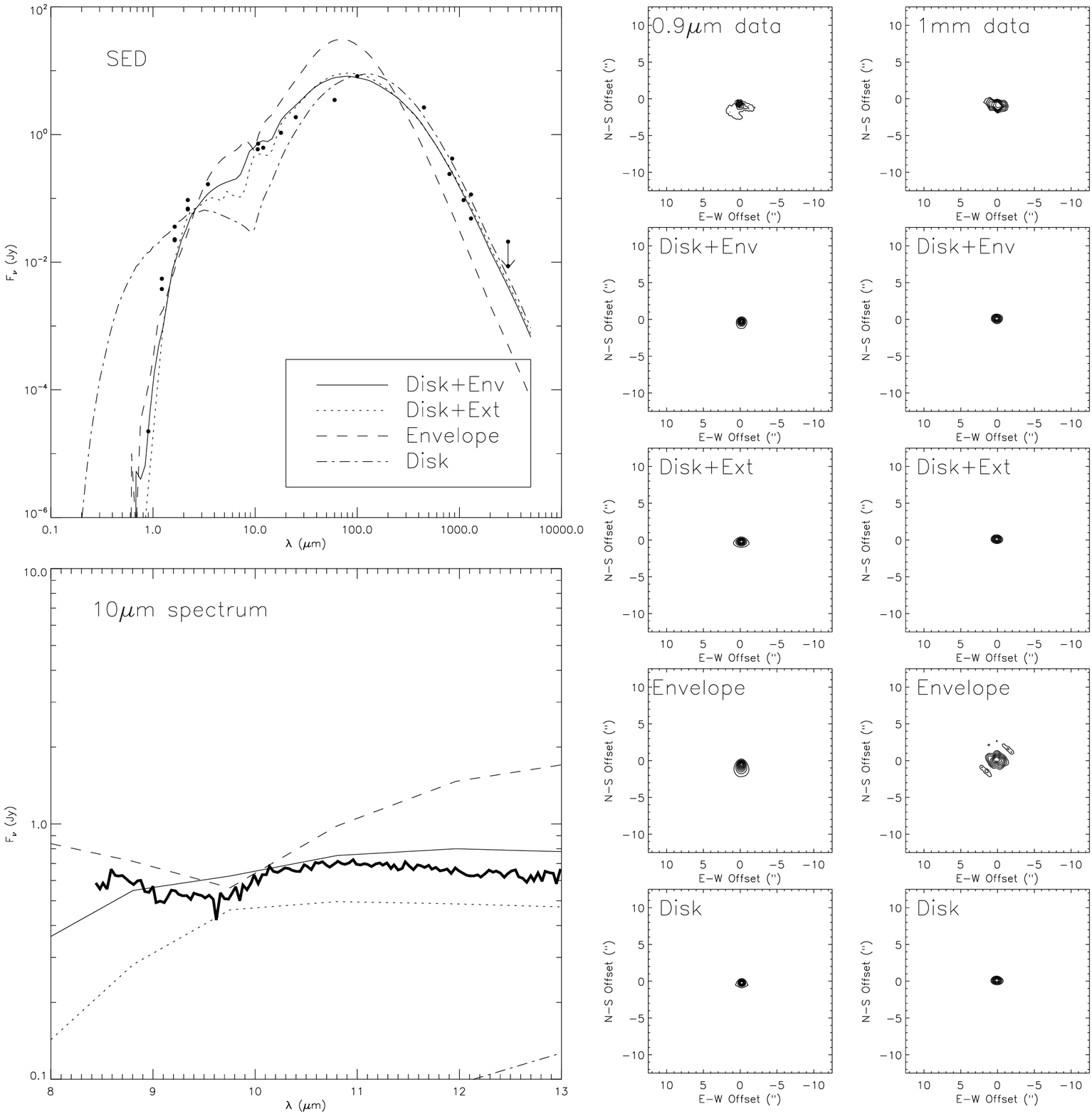}
\caption[Best-fit models for IRAS 04295+2251]
{SED data (Table \ref{tab:seds}), 10 $\mu$m spectrum 
\citep{KESSLER-SILACCI+05}, 0.9 $\mu$m scattered light image, 
and 1.3 mm continuum image for IRAS 04295+2251, along with best-fit models for
different assumed circumstellar geometries (Table \ref{tab:bestfits}).  
The observed images have been rotated on the sky by $155^{\circ}$ 
(east of north) so that they have the same position angle definition as the 
models.  Images have contour levels of 10\% of the peak flux,
beginning at 10\% ($\sim 13 \sigma$)
for the 0.9 $\mu$m images and 30\% ($\sim 5 \sigma$) for the 1 mm images.
Disk+envelope and disk+extinction models provide a good match to the data
for this source.  Note that the modeled 1 mm images are typically 
smaller than the observations in this case, since small values
of $R_{\rm c}$ are preferred by the SED and $I$-band data, forcing the
best-fit model to a smaller centrifugal radius, and thus a smaller 1 mm image
(see Figure \ref{fig:env_rc}).
\label{fig:i04295}}
\end{figure}

\begin{figure}
\plotone{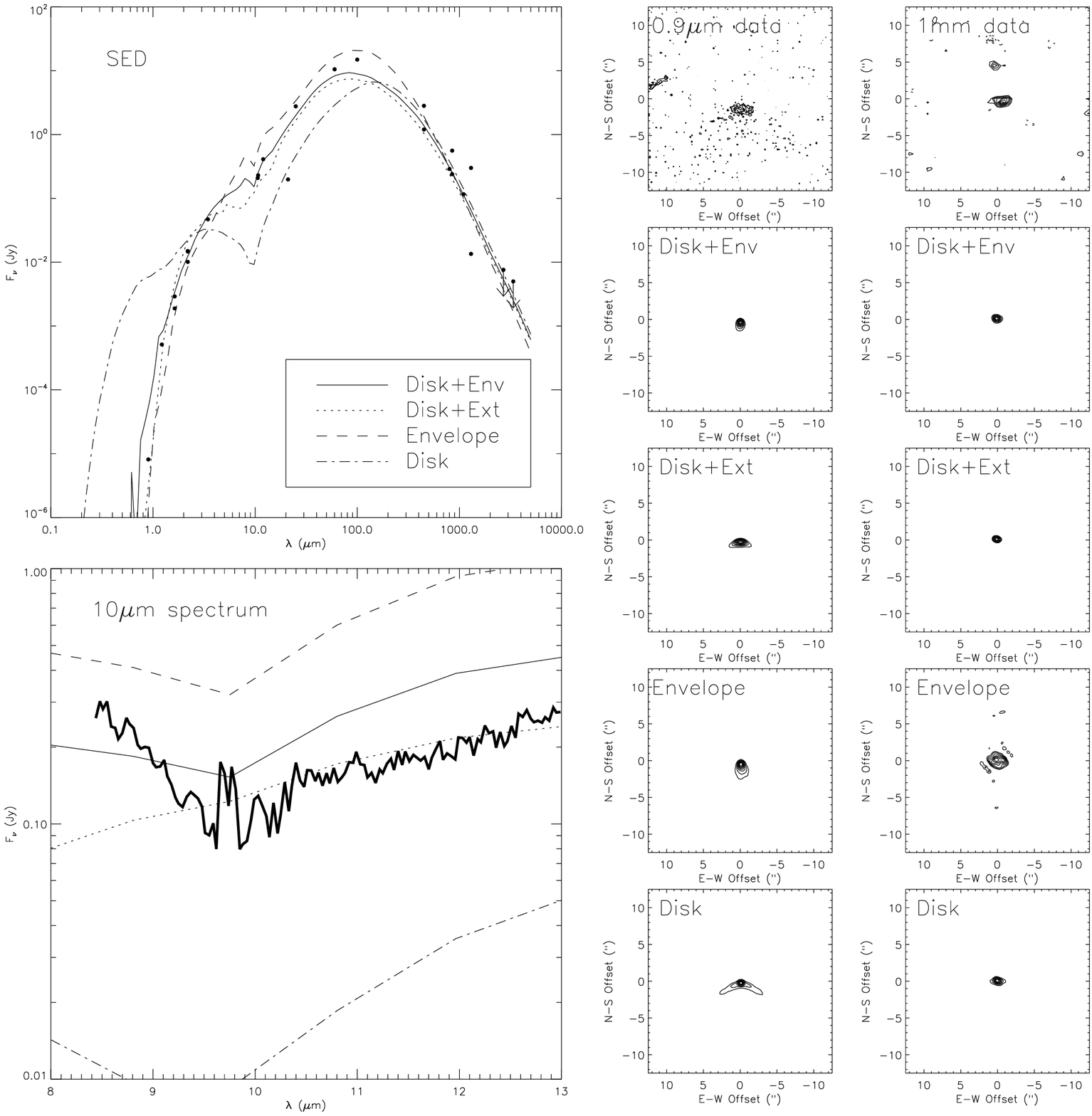}
\caption[Best-fit models for IRAS 04381+2540]
{SED data (Table \ref{tab:seds}), 10 $\mu$m spectrum 
\citep{KESSLER-SILACCI+05}, 0.9 $\mu$m scattered light image, 
and 1.3 mm continuum image for IRAS 04381+2540, along with best-fit models for
different assumed circumstellar geometries (Table \ref{tab:bestfits}).  
The observed images have been rotated on the sky by $180^{\circ}$ 
(east of north) so that they have the same position angle definition as the 
models.  Images have contour levels of 10\% of the peak flux,
beginning at 30\% ($\sim 2 \sigma$) for the 0.9 $\mu$m images and 
30\% ($\sim 2.5 \sigma$) for the 1 mm images. 
While pure disk models and pure envelope models do not provide an adequate 
match to the data, extincted disks and disk+envelopes do.
The disk+envelope and disk+extinction models provide the smallest residuals 
between models and data (Table \ref{tab:bestfits}). 
\label{fig:i04381}}
\end{figure}

For each of the density distributions described in \S \ref{sec:model},
we determine the model providing the best-fit to our combined SED+imaging
dataset.  
The properties of the best-fit models, as well as the reduced $\chi^2$ 
residuals between models and data, are listed in Table \ref{tab:bestfits},
and the models are plotted in Figures \ref{fig:i04016}--\ref{fig:i04381}.
Our results clearly indicate that pure disk models 
are not applicable for our sample of Class I
sources.  Rather, the data suggest 
models incorporating a massive envelope with an
outflow cavity, although 
most sources are fitted best by models including
{\it both} envelopes {\it and} disks.  In this section, we discuss
general results of our SED and image fitting, then describe results
for individual sources in detail in \S \ref{sec:results2}.

\begin{figure}
\plotone{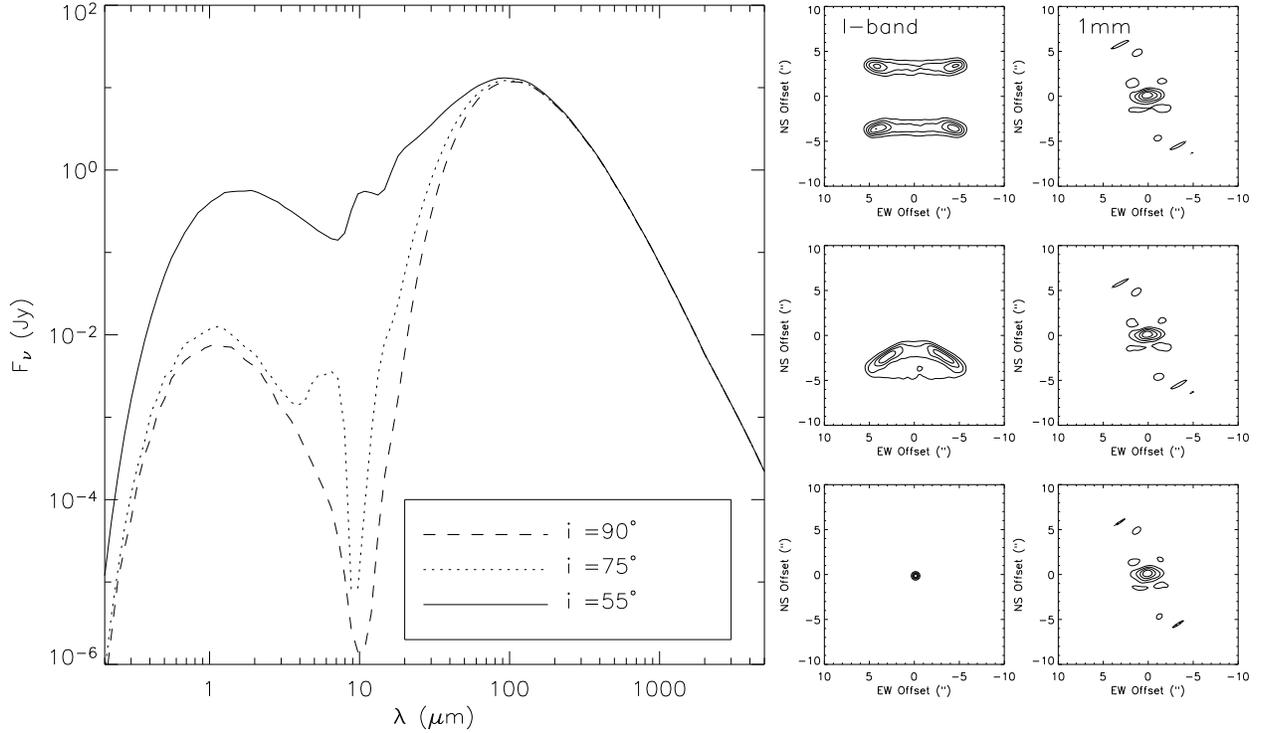}
\caption[Dependence of disk model on inclination]
{SEDs, scattered light images, and millimeter continuum images 
for a flared disk model at a range of viewing angles
($i$ increases from the bottom to top panels).  
This model assumes $M_{\rm disk}=5 \times 10^{-2}$ M$_{\odot}$,
$h_0=15$ AU, and $R_{\rm out}=1000$ AU.  The effects of inclination on disk
models are similar to those shown for envelope models in Figure 
\ref{fig:env_inc}.
More edge-on models exhibit deeper absorption at mid-IR wavelengths, and
higher extinction of the central star.  For moderate inclinations
($i \la 55^{\circ}$), the central star is visible, and dominates the
$I$-band emission.  For large inclinations, an asymmetric scattered light
structure is observed, while for nearly edge-on orientations, a symmetric,
double-lobed structure is observed.  The millimeter emission becomes
more extended in appearance as inclination
increases, although this is a minor effect.
\label{fig:disk_inc}}
\end{figure}

While the observed SEDs for our sample show fairly
shallow absorption at mid-IR wavelengths 
(Figures \ref{fig:i04016}--\ref{fig:i04381}), edge-on disk models
produce deep mid-IR absorption 
\citep[Figure \ref{fig:disk_inc}; see also][]{DALESSIO+99,WOOD+02,WPS03},
In addition, our asymmetric scattered light images rule out disk models
close to edge-on, since such models would produce symmetric structures 
(Figure \ref{fig:disk_inc}).  However, 
the extended scattered light images indicate that disk
models viewed close to face-on are not applicable either, since such models
would be dominated by emission from the point-like central protostar.  
Since we observe spatially resolved, asymmetric scattered light toward our 
sample objects, we know that we are not directly observing the star, as would 
be the case for face-on or moderately inclined disk models.  

Although there may be a narrow range of inclinations for which pure disk
models can fit both the SEDs and scattered light images of some of our
sources, substantial foreground extinction ($\ga 20$ mag) is needed
to make such a model consistent with the small optical/near-IR fluxes
for our sample. As indicated by Table \ref{tab:bestfits} and Figures
\ref{fig:i04016}--\ref{fig:i04381}, disk+extinction models provide
far superior fits than pure disk models with $A_V=0$. However, the
extinction values for the best-fit disk+extinction models are all 
much larger than expected from ambient material in the Taurus cloud
\citep[e.g.,][]{KH95}, indicating a large concentration of mass on small scales
toward these Class I sources.  Thus, the
disk+extinction models imply the existence of massive
envelopes in addition to disk components.  

Pure envelope models also fail to fit the data well for most targets in
our sample (Table \ref{tab:bestfits}).  As illustrated in Figures 
\ref{fig:i04016}--\ref{fig:i04381}, pure envelope models typically
over-predict the peak flux in the SED. As discussed further below,
pure envelopes also predict more extended millimeter emission than
actually observed.  Thus, pure envelope density 
distributions also do not seem suitable for explaining most
features of the data for
our sample.


For most, if not all, of the objects in our sample, the best fitting
models (i.e., those for which the residuals between model and data are 
minimized) incorporate both an envelope and a disk (described either
by disk+envelope or disk+extinction models; Table \ref{tab:bestfits}).  
However, while SEDs
and 1 mm images are fitted best by disk+envelope models, the scattered light 
images for some sources seem to favor pure disk or pure envelope models.  
This may suggest that our disk+envelope model is too simplistic, and that a 
more complex implementation of disk+envelope models would 
allow all of the data to be fitted simultaneously.  
Alternately, the dust grain size distribution or composition may differ from
our standard assumptions or vary radially or vertically within the model
density distributions.  One piece of evidence in support of the 
former hypothesis is the fact that the best-fit pure disk or envelope models, 
which usually fit the scattered light images well, generally
have larger values of $R_{\rm out}$ than the best-fit disk+envelope models
(Table \ref{tab:bestfits}).  This suggests a flattened
distribution of small dust grains that extends beyond the outer
radius of our best-fit disk+envelope models.  
We discuss this issue further in \S \ref{sec:geometry}.

One way to test the relative contributions of disks and envelopes is to
compare the compact millimeter emission observed in our OVRO observations
with more extended emission seen in lower-resolution single-dish
observations \citep{MA01}.  As illustrated in the case
of IRAS 04016+2610 in Figure \ref{fig:radprof}, pure disks, pure envelopes,
and disks+envelopes produce varying amounts of compact and extended emission,
and thus comparison of high- and low-resolution millimeter data can help
to distinguish between models.  As described
in \S \ref{sec:results2}, our interferometric fluxes typically (with one
exception) differ by $\sim 20-50\%$ from the
single-dish measurements of \citet{MA01} (Table
\ref{tab:compact}), indicating that the sources we
are observing produce most of their emission from compact inner regions.
Thus, disks appear to be an important, if not dominant, component of the
circumstellar distributions around our sample of Class I objects.  Combined
with the arguments presented above, this provides further evidence that the
circumstellar material resides in some combination of disks and envelopes.

\subsection{Results for Individual Sources \label{sec:results2}}
\subsubsection{IRAS 04016+2610 \label{sec:i04016}}
IRAS 04016+2610 is an IRAS source \citep{BEICHMAN+86} 
lying at the western edge of the
L1489 dark cloud \citep{BM89}, and driving a molecular outflow
\citep{MYERS+88,TVM89,MORIARTY-SCHIEVEN+92}.
Previous modeling of this source in terms of an infalling envelope model
found  $\dot{M} \sim 5  \times 10^{-6}$ M$_{\odot}$ 
yr$^{-1}$, $R_{\rm c} \sim 50$ AU, $i \sim 45-65^{\circ}$, 
and $L_{\ast} = 3.72$ L$_{\odot}$ \citep{KCH93,WKG97}.
An independent estimate of the inclination ($i \sim 60^{\circ}$)
was obtained from observations of a compact molecular outflow
\citep{HOGERHEIJDE+98}.
Previous observations of scattered light at near-IR wavelengths 
\citep{TAMURA+91,WKG97,PK02,ITI04} showed a similar morphology to our $I$ band
observations.   However, our observations enable for the first time accurate 
registration of the scattered light to the position of the central source, 
which allows better constraints on the circumstellar dust distribution than
possible in previous analysis.

Our best-fit model for IRAS 04016+2610 incorporates both a rotating, 
infalling envelope and an embedded disk.  This model fits all of our
data well, including the 0.9 $\mu$m image, 1 mm image, and SED (Figure
\ref{fig:i04016}).  The properties of this model are $\dot{M}=6 \times 10^{-6}$
M$_{\odot}$ yr$^{-1}$, $R_{\rm c}=100$ AU, $R_{\rm out}=2000$ AU, 
and $M_{\rm disk}=0.01$ M$_{\odot}$. The properties
of the envelope and disk components are consistent with previous analyses
in the context of either model individually \citep{KCH93,WKG97,HOGERHEIJDE01}.

In order to correctly fit the total luminosity of the system, we require
a large central luminosity, $L_{\rm central} \approx 4.7$ L$_{\odot}$.   This 
is substantially larger than the stellar luminosity inferred from near-IR
spectroscopy \citep{ITI04,WH04}, and suggests a large accretion luminosity.
Since the total accretion rate determined from our fitting of 
disk+envelope models is not substantially higher than for other sources,
we suggest that this additional luminosity may be generated in a more
active accretion shock near the protostellar surface.

For IRAS 04016+2610, the flux measured in our $\sim 2''$ beam is approximately
40\% of the value measured by \citet{MA01} in an $11''$ beam.  This is somewhat
larger than the ratios predicted by our best-fit models, and may indicate
that there is additional compact emission that we have not accounted for
in the models.  For example, if the disk is confined to radii smaller
than $R_{\rm c}$, as opposed to $R_{\rm out}$ as assumed in our models,
more mass would be concentrated at smaller radii.  Finally, we note that
although the pure envelope model appears to fit our combined imaging+SED
dataset almost as well as the disk+envelope model (Table \ref{tab:bestfits}),
the ratio of compact to extended emission for the pure envelope model is
much smaller than observed, providing further support for the disk+envelope 
model.

\subsubsection{IRAS 04108+2803B \label{sec:i04108}}
IRAS 04108+2803 is a $22''$-separation binary system 
\citep[e.g.,][]{DUCHENE+04} in the L1495 region,
and IRAS 04108+2803B is the component
that appears less environmentally evolved 
based on its spectral energy distribution; it 
emits the vast majority of the far-IR emission from the system. 
IRAS 04108+2803B also shows large scatter in photometric observations, 
indicating that it may be a variable star \citep{KCH93}.

Previous modeling of this object in the context of infalling
envelope models found $\dot{M} \sim 5 \times 10^{-6}$ M$_{\odot}$ 
yr$^{-1}$, $R_{\rm c} \sim 70-100$ AU, $i \sim 30-60^{\circ}$, and
$L_{\ast} = 0.63$ L$_{\odot}$. \citep{KCH93,WKG97}.
In contrast, \citet{CG99} showed that the SED of this object can be
fit well by a flared disk model with $i=65^{\circ}$, $\beta=1.2$, 
and $R_{\rm out}=270$ AU.  The small fractional polarization (5.1\%)
relative to other Class I sources \citep[$\sim 20\%$;][]{WKG97} was
invoked as further evidence that the scattered
emission from IRAS 04108+2803B arises in a disk rather than in the walls
of an outflow cavity \citep{CG99}.  Finally, IRAS 04108+2803B was
modeled as a T Tauri star in a disk that is dynamically warped by a 
hypothetical stellar companion \citep{TB96}.

Our results show that the data for this object cannot be fitted by 
a pure disk model, although the best fit is obtained for a model
incorporating a massive disk in addition to an
envelope (either disk+envelope of disk+extinction;
Table \ref{tab:bestfits}; Figure \ref{fig:i04108}).  
The best-fit disk+extinction model implies a disk with a mass of 0.6
M$_{\odot}$ surrounded by an envelope providing
25 mags of foreground extinction.
The best-fit disk+envelope model for this source
implies $\dot{M}=5 \times 10^{-6}$ M$_{\odot}$ yr$^{-1}$, $R_{\rm c}=30$ AU,
$R_{\rm out}$=500 AU, and $i=24^{\circ}$.
The central luminosity is close to what one would expect for a
T Tauri star, $L_{\rm central} \approx 0.4$ L$_{\odot}$.  The disk mass of 
our best-fit model is 0.5 M$_{\odot}$ (Table \ref{tab:bestfits}); however,
as discussed in \S \ref{sec:dmass}, this is likely an over-estimate.

The ratio of compact to extended millimeter emission (0.7;
Table \ref{tab:compact}) implies that a large fraction of the flux is
generated by a compact disk component. Thus, this source appears to be 
extremely disk-dominated, supporting the hypothesis of \citet{CG99}.  
On the other hand, a massive envelope component also appears necessary
to fit the imaging+SED data, 
consistent with previous models \citep[e.g.,][]{KCH93} and
recent {\it Spitzer} 
observations which attribute 15.2 $\mu$m CO$_2$ ice absorption
to a cold envelope \citep{WATSON+04}.

\subsubsection{IRAS 04239+2436 \label{sec:i04239}}
IRAS 04239+2436 is an IRAS source \citep{BEICHMAN+86} with a bright
scattered light nebula \citep{KENYON+93}, which has been tentatively
associated with high-velocity molecular gas \citep{MORIARTY-SCHIEVEN+92}.
Previous modeling of SEDs and near-IR scattered light (separately) in
the context of rotating, infalling envelopes, found
$\dot{M} \sim 2-4 \times 10^{-6}$ M$_{\odot}$ 
yr$^{-1}$, $R_{\rm c} \sim 10-70$ AU, $i=30-55^{\circ}$, and 
$L_{\ast} = 1.23$ L$_{\odot}$ \citep{KCH93,WKG97}.

Although the SED for this source is fit best by a disk+envelope model,
the combined dataset is fit comparably well by a disk+extinction model with
$A_V=20$ (Table \ref{tab:bestfits}; Figure \ref{fig:i04239}).  Thus, while 
there seem to be both disk and envelope components, the exact density profile 
of the source may not be matched exactly by either the disk+envelope or 
disk+extinction models.
As we discuss further in \S \ref{sec:geometry}, the fact that the
disk+envelope model fails to represent accurately our 0.9 $\mu$m image
may suggest that an additional flattened distribution of small dust grains
extending out to $\sim 2000$ AU
(not currently included in the model) is necessary to fit all of the
data simultaneously.  Assuming that our best-fit disk+envelope model provides
an accurate tracer of the mass, the properties of this source are
$\dot{M}=3 \times 10^{-6}$ M$_{\odot}$ 
yr$^{-1}$, $R_{\rm c} \sim 30$ AU, $R_{\rm out}=1000$ AU,
$M_{\rm disk}=0.5$ M$_{\odot}$,  $L_{\rm central} = 1.5$ L$_{\odot}$,
and $i=34^{\circ}$.  However, as discussed in \S \ref{sec:dmass}, 
this value of $M_{\rm disk}$ may be substantially over-estimated.

In addition to our modeling, there are several arguments suggesting that
IRAS 04239+2436 is surrounded by both a massive envelope and an embedded disk.
The ratio of compact to extended millimeter emission (Table \ref{tab:compact})
shows that a pure envelope model cannot fit the data, and thus there must
be a compact disk component.  On the other hand, recent 
{\it Spitzer} observations 
find a deep 15.2 $\mu$m CO$_2$ ice absorption feature, which likely arises
in a large, cold envelope \citep{WATSON+04}.  These arguments provide
further support for a combined disk+envelope distribution of material
around this source, and motivate future modeling that can fit all of the
data simultaneously.

\subsubsection{IRAS 04295+2251 \label{sec:i04295}}
IRAS 04295+2251 has appeared point-like in previous near-IR imaging
observations \citep[e.g.,][]{PK02}, and is tentatively
associated with a molecular outflow \citep{MORIARTY-SCHIEVEN+92}.
Previous investigators analyzed SEDs and scattered light
images (separately) in the context of infalling envelope models, and
found $\dot{M}=1-5 \times 10^{-6}$ M$_{\odot}$ 
yr$^{-1}$, $R_{\rm c} = 70-100$ AU, $i=0-30^{\circ}$, and
$L_{\ast} = 0.44$ L$_{\odot}$ \citep{KCH93,WKG97}.
It has also been suggested that a pure disk model may be able to
fit the SED for this source \citep{CG99}.

Our modeling shows that the best-fit is obtained for a density distribution
including both envelope and disk components (Table \ref{tab:bestfits};
Figure \ref{fig:i04295}).  Formally, the best-fit is obtained for
a disk+extinction model including a 1.0 $M_{\odot}$ disk surrounded by
an envelope that provides 20 magnitudes of extinction.  The disk+envelope
model also provides a good fit to the data, and the properties of this model
are $\dot{M}=4 \times 10^{-6}$ M$_{\odot}$ 
yr$^{-1}$, $R_{\rm c} \sim 100$ AU, $R_{\rm out}=500$ AU,
$M_{\rm disk}=1.0$ M$_{\odot}$, $L_{\rm central}=0.6$ L$_{\odot}$,
and $i=22^{\circ}$.
As for other sources in our sample, the inferred values of $M_{\rm disk}$ 
for best-fit models are probably too high (\S \ref{sec:dmass}).
The ratio of compact to extended millimeter emission observed
for IRAS 04295+2251 (0.42; Table \ref{tab:compact}) is consistent
with that predicted for the best-fit disk+envelope model, and implies
a relatively disk-dominated density distribution.

\subsubsection{IRAS 04381+2540 \label{sec:i04381}}
IRAS 04381+2540 is in the B14 region, and is associated with high-velocity
molecular gas \citep{TVM89}.  In addition, extended near-IR emission
has been observed toward this object \citep{TAMURA+91}.
Previous modeling in the context of a rotating, infalling envelope found
$\dot{M}=5-13 \times 10^{-6}$ M$_{\odot}$ 
yr$^{-1}$, $R_{\rm c} = 50-300$ AU, $i=30-75^{\circ}$, and
$L_{\ast} = 0.66$ L$_{\odot}$ \citep{KCH93,WKG97}.
Moreover, the inclination was estimated independently from observations of a 
compact molecular outflow to be $\sim 40-70^{\circ}$ \citep{CHANDLER+96}.

Our modeling shows that the best fit is obtained for a density distribution
incorporating both an envelope and a disk (Table \ref{tab:bestfits};
Figure \ref{fig:i04381}). The properties of the best-fit
disk+envelope model are $\dot{M}=9 \times 10^{-6}$ M$_{\odot}$ 
yr$^{-1}$, $R_{\rm c} = 30$ AU, $M_{\rm disk}=1.0$ M$_{\odot}$,
$i=38^{\circ}$, and $L_{\rm central}=0.6$ L$_{\odot}$.  
The disk+extinction model also provides a good 
fit, and may provide a more accurate representation of the 0.9 $\mu$m
scattered light image.   These results indicate that the circumstellar
dust distribution around IRAS 04381+2540 is dominated by a disk, although
a massive envelope is also required to fit the data. However, the values of 
$M_{\rm disk}$ in Table \ref{tab:bestfits} are probably
over-estimated (\S \ref{sec:dmass}) and the disk mass may be closer to
$\sim 0.2$ M$_{\odot}$ for this object.

A potential problem with the best-fit disk+envelope model is that it does
not correctly reproduce the observed ratio of compact to extended millimeter
flux (Table \ref{tab:compact}).  Rather, this ratio indicates that the
density distribution for this source is envelope-dominated.
This may indicate that there is envelope material distributed uniformly
over large spatial scales, leading to large amounts of extended flux being
resolved out in our OVRO image.  This is consistent with an extended structure
extending to large radii from the central source ($\ga 3000$ AU) observed
by \citet{YOUNG+03}, and may favor the disk+extinction model, which allows 
a large mass of dust to be distributed out to larger radii.
Regardless of the exact density distribution, it seems that both disk and
envelope components are necessary to fit the data for this source.

The high derived accretion rate for IRAS 04381+2540, as well as the
strong indication from Table \ref{tab:compact} that the source is 
envelope-dominated, suggest that this object may be evolutionarily younger
than other sources in our sample.  Moreover, the mass accretion rate
may be underestimated for this source: the mass of this object has been 
estimated dynamically from millimeter spectral line observations to be 
$\sim 0.2-0.4$ M$_{\odot}$ \citep{BC99}, somewhat lower than our assumed value 
of 0.5 M$_{\odot}$, which implies that the value of $\dot{M}$ derived above 
may be underestimated by $\sim 10-30\%$ (Equation \ref{eq:env}).

\section{DISCUSSION \label{sec:disc}}
In \S \ref{sec:results}, we fit our combined imaging+SED data for a sample
of Class I sources with several models for the circumstellar dust distribution:
flared disks, collapsing envelopes, and combinations of disks and envelopes.
Of the models considered, disk+envelope (and/or disk+extinction)
density distributions generally
provide the smallest residuals between models and data 
(Table \ref{tab:bestfits}).  The properties of best-fit 
models are similar for the five sources in our sample, and for most
parameters the spread in best-fit values is less than an order of 
magnitude.  These tightly-clustered values are not 
surprising given the selection criteria for our sample and
the similarity of the observational data for our targets
(\S \ref{sec:sample}; Figures
\ref{fig:i04016}-\ref{fig:i04381}).  However, these similar parameter values
may also be due in part to finite parameter sampling and
limitations in our models.

In this section, we discuss potential modifications to our disk+envelope 
models that may be required to provide better fits to combined imaging+SED 
data, and examine the physical plausibility of derived model parameters.
We also use our results to understand better the evolutionary stage of 
Class I sources, and to place them in context relative to the better-studied 
Class II objects.  Finally, we discuss how
new astronomical instruments will improve constraints on the circumstellar
dust distributions for Class I objects.


\subsection{Large-Scale Geometry \label{sec:geometry}}
While previous investigations of Class I sources argued for either pure
envelope or pure disk models \citep[e.g.,][]{KCH93,WKG97,CG99}, our results
based on simultaneous modeling of scattered
light images, thermal images, and SEDs  (\S \ref{sec:results})
indicate that the most suitable dust
distributions likely incorporate massive envelopes {\it and} 
massive embedded disks. Given the large envelope centrifugal radii determined
for our sample, the fact that massive disk components are also required is
not surprising.  In models of rotating infalling envelopes, 
the centrifugal radius demarcates the point at which infalling material
piles up due to conservation of angular momentum
\citep[e.g.,][]{TSC84,KCH93}.  Because the centrifugal radius grows with
time as the collapsing cloud rotates faster 
\citep[$R_{\rm c} \propto t^3$; e.g.,][]{HARTMANN98},
material should have previously piled up within 
the current value of $R_{\rm c}$, creating
a dense disk.  Moreover, viscous spreading tends to smear out the
material piled up at $R_{\rm c}$ into a more disk-like distribution.
Since this material is not accounted for in 
the envelope density distribution (Equation \ref{eq:env}), it is not
surprising that the addition of a disk component improves the agreement
between model and data.

Although models incorporating both disks and envelopes 
generally yield the best fits to our
combined SED+imaging dataset (Table \ref{tab:bestfits}), the exact
density distribution is not firmly constrained since disk+envelope and
disk+extinction models often provide fits of comparable quality.  
In addition, the scattered light images for some sources are fit better 
by pure disk or envelope models with large ($\sim 2000$ AU) outer radii.  
Since scattered light probes trace material in surface layers,
our 0.9 $\mu$m images may contain contributions from flattened distributions 
of tenuous material at larger radii.
Moreover, our model assumes that the dust grain
properties are the same everywhere.  If one takes into account the effects of 
dust settling \citep[e.g.,][]{GW73}, 
then the grains in the disk surface layer and outer 
envelope may be smaller than those in the dense midplane, allowing more 
scattering for a given mass of dust \citep[e.g.,][]{WPS03,WHITNEY+03b}.

The large polarizations observed toward Class I sources in near-IR 
polarimetric imaging observations support the notion of extended
distributions of small dust grains \citep{WKG97,LR98}.  At the edges of
near-IR scattered light nebulosity observed for our sample,
where most of the light is single-scattered at approximately 90$^{\circ}$,
linear polarizations are 70-80\% \citep{WKG97}.  In contrast, the maximum
polarizations predicted\footnote{The MC3D 
code has the capability of computing scattered
light images in different Stokes parameters, allowing recovery of
polarimetric information \citep{WOLF03}.}  
by our disk+envelope $I$-band model images 
are $\sim 25\%$.  Moreover, the integrated polarizations in our model images
are substantially lower than the integrated near-IR polarizations measured
by \citet{WKG97}, 
re-enforcing our hypothesis that there may be a population of small grains
at large radii not included in our models.

Since our best-fit models indicate that Class I sources are probably 
surrounded by envelopes with outflow cavities (in addition to disks), we 
expect observed outflows from these objects to lie in the middle of these 
cavities.  For IRAS 04016+2610, IRAS 04239+2436, and IRAS 04381+2540,
outflows have been observed at position angles of 165$^{\circ}$,
$45-60^{\circ}$, and $10^{\circ}$, respectively 
\citep{HOGERHEIJDE+98,GWK97,SAITO+01}.  Comparison of
these position angles with Figure \ref{fig:iband} demonstrates that the
outflows lie in the middle of the observed scattered light structures.  
For IRAS 04295+2251 and IRAS 04108+2803B, no
geometrical information about molecular outflows is available.
Thus, in all cases where outflow geometries can be derived, they are
consistent with expectations from our modeling.

The range of inclinations for best-fit models is small, with a spread of
$\la 20^{\circ}$ across all sources for a given model.  This may be explained
to some extent by selection effects: bright scattered light can only be
observed if targets are inclined sufficiently to block the light from the
central star.  Moreover, scattered light is somewhat brighter for
moderate than for edge-on inclinations, which may bias our sample against
edge-on sources as well.  Nevertheless, the tight clustering
of inclinations (Table \ref{tab:bestfits}) may be due in some part to 
limitations of 
our models in accurately representing the true density distributions. 
For example, for the disk+envelope model the
assumed disk density profile may favor small inclinations for which
the observer's line of sight passes through the flared surface of the disk 
rather than the dense midplane, while inclinations $\ga 30^{\circ}$, for
which the line of sight does not pierce directly down the outflow cavity, are
required to reproduce the extended scattered light structures and heavily
reddened SEDs.  Observations of larger samples and refinements to the models 
are necessary to provide a more reliable estimate of the inclination
distribution of Class I sources.

\subsection{Disk and Envelope Masses \label{sec:dmass}}
\begin{figure}
\plotone{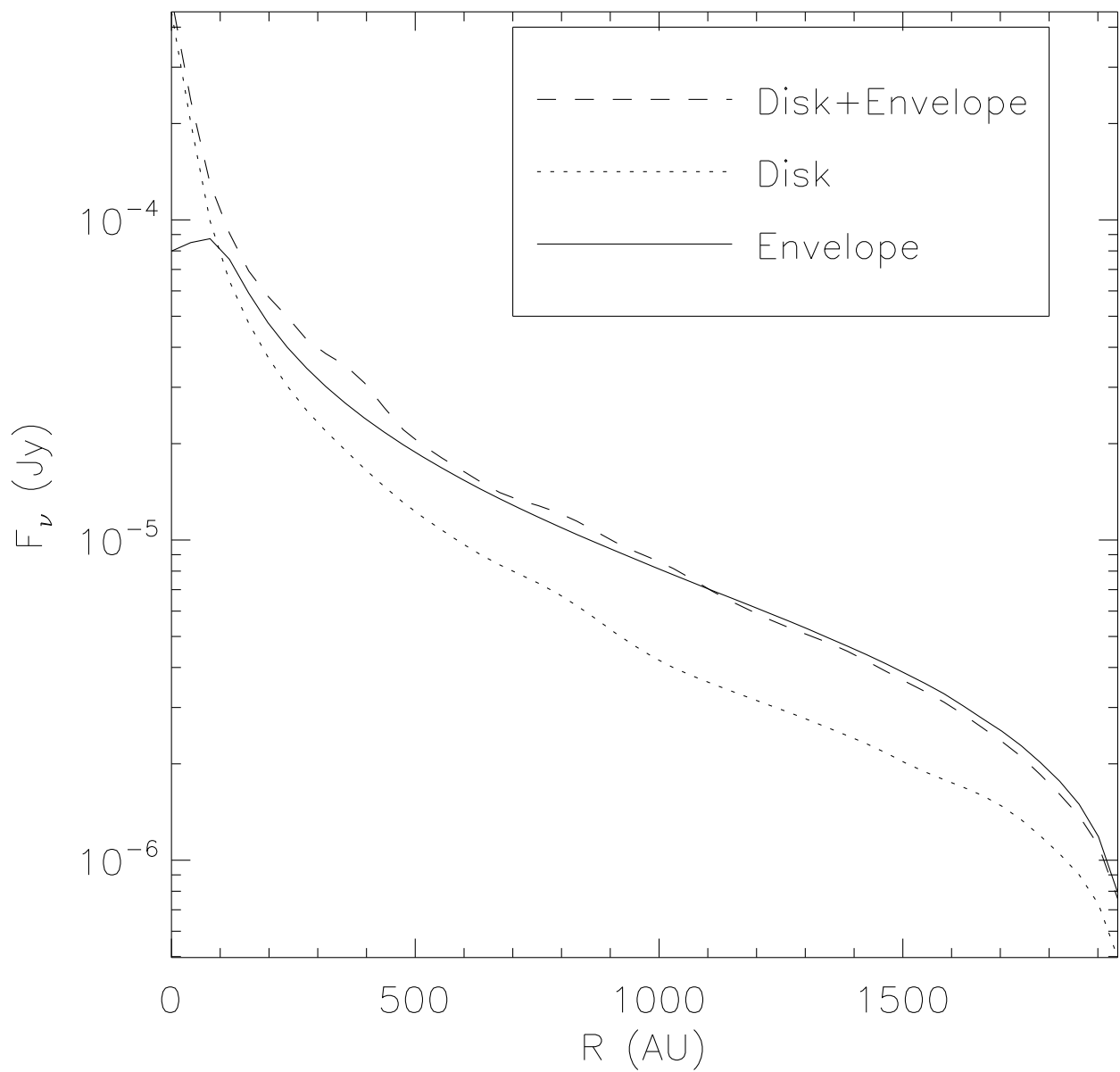}
\caption[Radial profiles of $\lambda$1.3 mm emission of best-fit models for 
IRAS 04016+2610]{Azimuthally-averaged radial profile of millimeter emission of 
best-fit models for IRAS 04016+2610 (Table \ref{tab:bestfits}).  
The pure envelope model (solid line) 
produces most of its millimeter emission near the centrifugal radius, 100 AU
in this case.  The pure disk model (dotted line) is more centrally-peaked,
producing the majority of the millimeter emission at small radii. The
disk+envelope model resembles the pure disk at small radii and the 
pure envelope at large radii. 
\label{fig:radprof}}
\end{figure}

As seen in Table \ref{tab:bestfits}, the masses of best-fit disk models,
and the masses of the disk components of disk+envelope models, span 
a range of values from $\sim 0.01$ to 1.0 M$_{\odot}$.  These high disk
masses result naturally from the high optical depths of these models
($\tau > 10^4-10^6$ at 100 $\mu$m, and $>10-1000$ even at 1 mm).
For pure disk models, large masses are required to extinct the
central star (thus allowing scattered light to be observed) and to
produce the observed millimeter emission (Figure \ref{fig:disk_mass}).
Moreover, the high optical depths that result from large disk masses
lead to large amounts of cold dust in the midplane, which emits most
of its radiation at millimeter wavelengths.  For disk+envelope models, 
large disk masses enable enhanced millimeter emission without
substantially altering the peak flux at shorter wavelengths.  Moreover,
substantial disk components lead to more centrally-concentrated millimeter
emission, consistent with observations 
(Figure \ref{fig:radprof}; Table \ref{tab:compact}).

Disk masses larger than $\sim 0.1$ M$_{\odot}$ may be unphysical, since
they are gravitationally unstable \citep[e.g.,][]{LB94}.
Specifically, when the mass of a rotating disk is larger than approximately
$M_{\ast} H/R$ (where $H/R$ is the
disk aspect ratio), the disk becomes gravitationally unstable and
rapidly transfers angular momentum outward, resulting in rapid accretion
of material onto the central protostar.  This accretion process occurs on
the order of the outer disk dynamical timescale ($\sim 10^3-10^4$ yr), and
is thus very fast compared to the inferred infall rate and should rapidly
bring the disk mass down to a stable level.  For our best-fit disk models,
$H/R \sim 1/10$, and for a 0.5 M$_{\odot}$ star the stability
criterion requires $M_{\rm disk} \la 0.05$ M$_{\odot}$.  While numerical
simulations suggest that disks may remain stable with slightly
higher masses, $\sim 0.3 M_{\ast} \approx 0.15$ M$_{\odot}$ 
\citep[e.g.,][]{LB94,YBL95}, our models still 
require disk masses far in excess of this value.

One implication of these un-physically-large disk masses is that rotating
disk models seem untenable for modeling the imaging and SED data for
Class I sources. Our massive, best-fit disk models may therefore resemble the
magnetically-supported ``pseudo-disks'' proposed to occur during the early 
stages of protostellar collapse \citep{GS93a,GS93b}. Thus, while 
disk+extinction models provide good fits to the data for some sources
(\S \ref{sec:results}; Table \ref{tab:bestfits}), the large disks implied
by these models may not correspond to the Keplerian disks seen
at later evolutionary stages.

These large disk masses are also a concern for the best-fit disk+envelope
models, where disk components are necessary to fit the SEDs well
(Figures \ref{fig:i04016}-\ref{fig:i04381}),
and to provide the correct ratios of compact to extended millimeter emission
(Table \ref{tab:compact}).  As illustrated in Figure \ref{fig:radprof},
the most important contribution of the disk component occurs within 
the envelope's centrifugal radius, while outside of $R_{\rm c}$, the
millimeter emission from a disk+envelope model has a similar profile to
the emission from a pure envelope.  If the disk
component is truncated at $R_{\rm c}$ (as opposed to $R_{\rm out}$, as
implemented in our radiative transfer code), the disk mass can be
decreased by a factor of $\sim 5$, while the compact disk emission
within $R_{\rm c}$ will still produce ratios of compact to extended
millimeter emission consistent with observations.
In order to produce the correct total millimeter flux, the envelope
mass must be increased to compensate for the loss of emission from 
larger disk radii.  As indicated by Figure \ref{fig:radprof}, these outer
disk regions contribute approximately 50\% of the millimeter flux at radii
$\ga R_{\rm c}$.  By truncating the disk component of disk+envelope 
models at $R_{\rm c}$, the disk masses listed in Table \ref{tab:bestfits}
may be decreased by a factor of $\sim 5$, while the envelope masses may
increase by factors of $\sim 2$.  Thus, true masses of the disk components
for disk+envelope models may be $\la 0.2$ M$_{\odot}$, close to the limit of 
gravitational stability and therefore more physically plausible than the 
values listed in Table \ref{tab:bestfits}.

Even for disks truncated at $R_{\rm c}$, the disk masses are often larger
than the masses of the envelope components of disk+envelope models.
Envelope masses range from $5 \times 10^{-3}$ to 0.05 M$_{\odot}$, and
sources with smaller envelope masses tend to have relatively
large disk masses, suggesting that different objects in our sample may be
more or less disk-dominated.  However, large-scale emission from extended
material belies this trend to some extent: for example, while IRAS
04381+2540 appears to be the most disk-dominated source in our sample
based on estimated disk and envelope masses, Table \ref{tab:compact}
indicates that there is probably a substantial extended dust component as well.
Thus, it is hard to draw conclusions about the relative evolutionary stages
of different objects in our sample based on inferred disk and envelope masses.

\subsection{Evolutionary Stage \label{sec:evolution}}
An issue closely related to the properties of circumstellar material in Class I
sources is their evolutionary stage.  While flared disk models are consistent
with the SEDs of Class II objects 
\citep[e.g.,][]{KH87,CG99,DDN01,EISNER+04,LEINERT+04},
we have shown that such models are not suitable for the
Class I objects in our sample.  Indeed, we have found that models incorporating
both envelopes and disks provide a better match to the data.  This finding
would seem to support
the standard assumption that Class I sources are more
embedded, still surrounded by massive envelopes, and potentially
younger than Class II sources.  On the other
hand, spherically symmetric models \citep[e.g.,][]{LARSON69,SHU77} are not
compatible with our data, indicating that Class I sources are at a
stage intermediate to cloud cores and star+disk systems.

However, other investigators have suggested that Class I and II sources
actually may be at similar evolutionary stages. One piece of evidence in
support of this hypothesis is that the spectroscopically-determined stellar 
ages of Class I and II sources appear indistinguishable \citep{WH04}.
It has also been suggested that differences in observed circumstellar 
properties between the two classes are caused by different viewing angles,
rather than different amounts or geometry of circumstellar material
\citep{CG99,WH04}.  However, our results show that the five Class I
objects in our sample are viewed at moderate inclinations 
($\sim 30-50^{\circ}$), refuting earlier suggestions that
optically-visible Class I objects are predominantly viewed edge-on,
and arguing that the circumstellar material around Class Is is indeed
less evolved than that around Class IIs. 

Another argument supporting similar evolutionary states of Class I and II
objects is the fact that measured accretion rates, pertaining to the transfer
of material from the inner disk onto the central star, appear similar for
the two classes \citep{WH04}.  In contrast, 
the new data and modeling presented in
this paper confirm earlier estimates of mass infall rates from the envelope
onto the disk orders of magnitude higher than the derived inner disk
accretion rates.
The fact that disk and envelope accretion rates are not the same suggests
periodic ``FU-Ori'' episodes where the accretion rate temporarily increases
by more than an order of magnitude, likely due to a
gravitational instability in the 
accretion disk \citep[e.g.,][and references therein]{BL94,HK96}.

As noted by \citet{WH04}, the episodic accretion 
scenario would imply more massive
disks in Class I sources relative to Class II objects.  While Class I
objects appear to be surrounded by larger total masses of dust, the
emission on scales smaller than $\sim 12''$ indicates similar
masses in the compact disk components \citep{AM94,MA01,WH04}.  However, 
the conversion of millimeter flux into mass depends on the optical depth,
dust opacity and temperature;  the higher optical depths and cooler dust
temperatures in Class I sources may therefore lead to higher masses.
Estimated disk masses for our best-fit disk+envelope models 
(\S \ref{sec:dmass}) are larger than typically observed
for Class II objects and are close to the values required for gravitational
instability, providing support for the non-stationary accretion model.

\subsection{Further Constraints on the Horizon \label{sec:future}}

We have demonstrated the power of combining spatially-resolved images
at multiple wavelengths with broadband spectral energy distributions  
when modeling the dust distributions around Class I sources.
Low-density surface layers, hot inner regions, and cool outer
regions are traced by optical/infrared scattered light, 
thermal mid-IR emission, and far-IR and millimeter emission, respectively.
Because images at these various wavelengths probe different regions 
of the circumstellar material, and even different emission mechanisms, 
they place tight constraints on the range of circumstellar
dust models consistent with the data, especially when used in conjunction.
In this section we discuss the near-future prospects for spatially
resolved imaging of Class I sources.

Although our marginally-resolved millimeter images place important
constraints on the centrifugal and outer radii (and viewing angle when
combined with the position of near-IR scattered light), well-resolved
images will provide better inclination constraints and 
useful additional information including direct measurement 
of the radial intensity profile and, in the best cases, the 
vertical intensity profile; taken together these constrain the density and
temperature profiles.  Analysis of spectral lines with different excitation
conditions can probe directly the density and temperature structure.
The enhanced angular resolutions and sensitivities of new and upcoming
millimeter interferometers, including the SMA, CARMA, and ALMA, will
enable detailed studies of Class I sources.

At shorter wavelengths, the FWHM angular sizes of our typical disk+envelope
model images (at 140 pc) are  $\sim 0\rlap{.}''3$ to $1\rlap{.}''2$
at 70 to 160 $\mu$m.  Though smaller than the beam size of {\it Spitzer},
future far-IR interferometers such as the mission concepts {\it  
SPIRIT} and {\it SPECS} will resolve this thermal emission.
At 10-20 $\mu$m the emission is generated at small radii ($\la 10$ AU), 
tracing dense material close to the star which can be studied
with ground-based mid-IR interferometry.  
At 10 and 18 $\mu$m, the FWHM sizes are $\sim 30$ and 40 mas, and the 
1\% emission contours lie at 140 and 280 mas, though
at low flux levels ($\la 30$ mJy).
Current sparse-aperture interferometry with the Keck telescope
can achieve angular resolutions of $\sim 100$ mas for sources  
brighter than a
few Jy \citep[e.g.,][]{MONNIER+04}, while future instruments
like LBTI should yield additional improvements in resolution and  
sensitivity.
Future capabilities such as mid-infrared instrumentation on TMT
may also be able to achieve the required resolution and sensitivity.
We note that different density distributions (e.g., pure envelopes
or pure disks) can produce less centrally-peaked emission, which would
lead to larger and thus more easily resolved images at infrared
wavelengths.

Finally, spectroscopy from the {\it Spitzer Space Telescope}
will add valuable constraints on dust
mineralogy and particle size distribution for Class I sources
\citep[e.g.,][]{WATSON+04}.
Moreover, the shape of spectral
features can provide a sensitive probe of dust optical depth and source
inclination.  Since different
molecules, ices, and dust species arise in different physical  
conditions,
{\it Spitzer} observations will thus also provide powerful constrains on
large-scale geometry.

\section{Conclusions \label{sec:conc}}
We imaged a sample of five embedded Class I sources in the Taurus star
forming region in 0.9 $\mu$m scattered light and thermal 1 mm continuum 
emission, and we analyzed these data together with spectral energy 
distributions and 10 $\mu$m spectra from the literature.  Using the MC3D
Monte Carlo radiative transfer code, we generated
synthetic images and SEDs for four classes of models: 1) rotating infalling
envelopes including outflow cavities; 2) flared disks; 3) disks+envelopes;
and 4) disks+extinction.  For each class of model, we sampled small grids
of relevant geometric parameters and determined the circumstellar dust 
distributions providing the best fits to our data in a $\chi^2$ sense.

The imaging and SED data are generally inconsistent with either pure disk
or pure envelope models, and we find that the best fits are obtained
with models incorporating both massive envelopes and massive embedded disks.
Given the large centrifugal radii derived for our sample ($R_{\rm c} = 30-100$
AU), the need for massive disks is not necessarily surprising.  While not
included in the rotating infalling envelope model, one expects a dense disk
of material interior to $R_{\rm c}$ because of the growth of centrifugal 
radius with time: at earlier times, $R_{\rm c}$ was smaller, and thus
material should have gradually piled up at successively larger radii
out to the current value of $R_{\rm c}$.  Thus, models incorporating
both envelopes and disks may be the most accurate (of those considered)
for representing physical infalling envelopes.

However, our results indicate that refinements to the models are necessary.
For example, disk+envelope models including disk components truncated
at $R_{\rm c}$ may yield disk masses closer to physically plausible values.
In addition, our scattered light images point to the existence of
extended, tenuous distributions of material, which must be included in
the models in order to fit all of the data simultaneously.  As more
spatially-resolved data is collected over a broad range of wavelengths, 
further refinements to the models will likely be warranted.

The overall geometry inferred for our sample of Class I objects is neither
spherically-symmetric, as expected for the earliest stages of cloud collapse,
nor completely flattened as seen in Class II sources.  Thus, our models 
confirm the picture where Class I sources are at an
evolutionary stage intermediate to collapsing cores and fully assembled
stars surrounded by disks.

The mass infall rates derived for our sample are between $3 -9 \times 10^{-6}$
M$_{\odot}$ yr$^{-1}$, consistent with previous results. 
However, these infall rates
are more than an order of magnitude higher than accretion rates pertaining
to the transfer of material from the disk onto the central star derived
from high-dispersion spectroscopy \citep{WH04}.  
This discrepancy argues for periodic
``FU Ori'' episodes of increased accretion:  since infalling material
is not accreted onto the star at the same rate, material piles up in the disk.
Once the disk becomes unstable to gravitational instability, the accretion
rate rises dramatically for a short period, depleting the disk mass and
restoring stability.  The ``FU Ori'' accretion hypothesis is also
consistent with the high disk masses of disk+envelope models, which suggest
disks near to the limit of gravitational stability.

We have demonstrated the power of modeling broadband SEDs in conjunction with
images at multiple wavelengths.  In addition, 
we showed that 10 $\mu$m spectra can provide valuable additional constraints
on circumstellar geometry, since the depth and shape of spectral features
depends on dust optical depth and source inclination.  Future observations
with upcoming millimeter interferometers, the {\it Spitzer Space Telescope},
and other instruments will provide additional spatially- and
spectrally-resolved information, greatly enhancing constraints on the
circumstellar dust around Class I objects.

\medskip
\noindent{\bf Acknowledgments.}
The 0.9 $\mu$m images and 18 $\mu$m photometry presented in this paper
were obtained at the W.M. Keck Observatory, which is 
operated as a scientific partnership among California Institute of Technology,
the University of California, and NASA.  The Observatory was made possible
by the generous financial support of the W.M. Keck Foundation.
The authors wish to recognize and acknowledge the cultural role and reverence 
that the summit of Mauna Kea has always had within the indigenous Hawaiian 
community. We are most fortunate to have the opportunity to conduct 
observations from this mountain.  
This publication makes use of data products from the Two Micron
All Sky Survey, which is a joint project of the University of Massachusetts
and the Infrared Processing and Analysis Center, funded by the National
Aeronautics and Space Administration and the National Science Foundation.
2MASS science data and information services were provided by the Infrared
Science Archive at IPAC.  The authors wish to thank R. Sari and R. White
for stimulating and constructive conversations about this work.
J.A.E. acknowledges support from a Michelson Graduate Research Fellowship,
and J.M.C. acknowledges support from 
the Owens Valley Radio Observatory, which is supported by the National 
Science Foundation through grant AST 02-28955.  S.W. was supported by the 
German Research Foundation (DFG) through the Emmy Noether grant WO 857/2-1.

\appendix
\section{LRIS Images of Larger Sample \label{sec:app_lris}}
\begin{figure}
\plotone{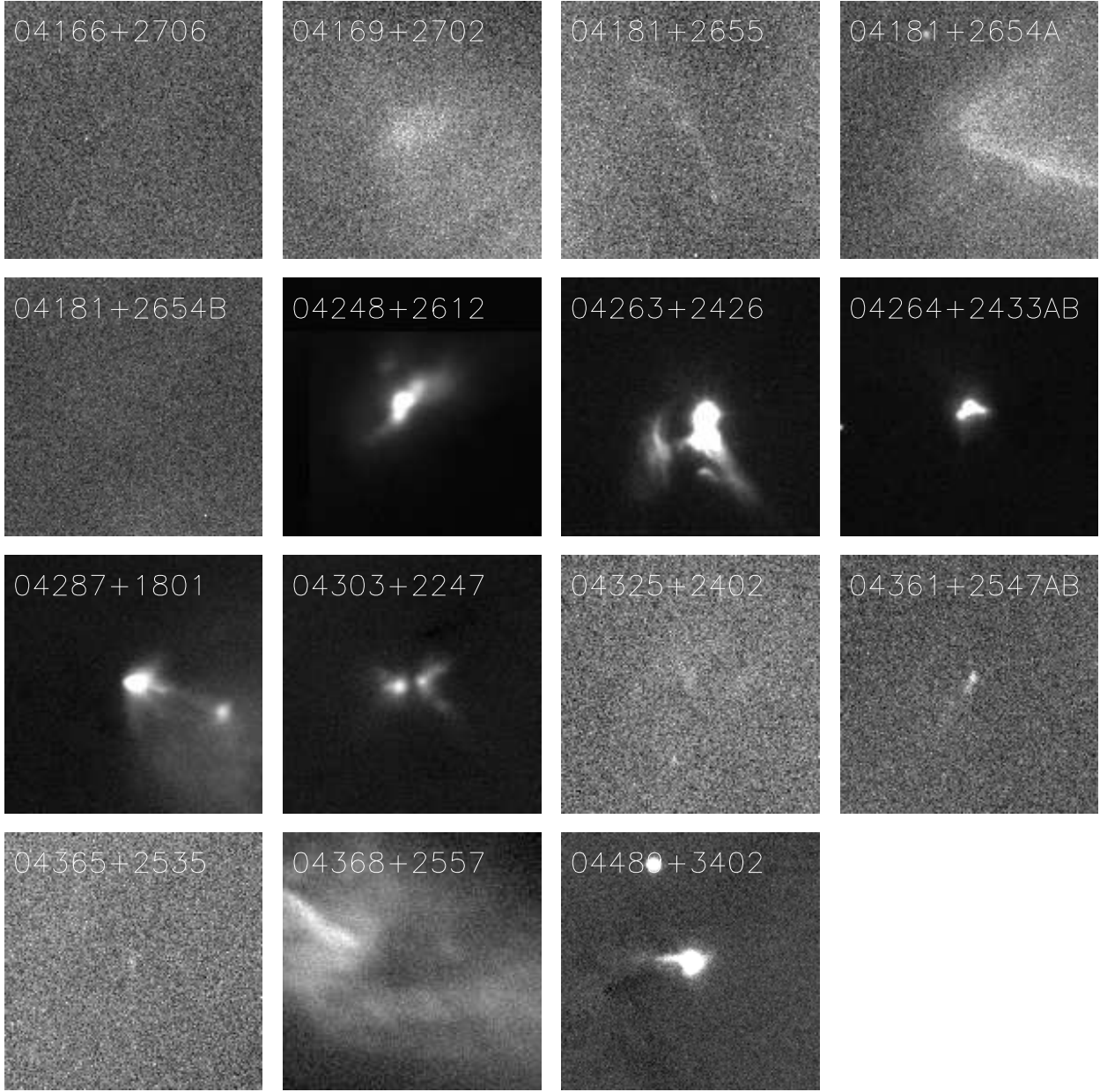}
\caption[$\lambda$0.9 $\mu$m scattered light images of larger sample]
{Keck/LRIS images of the Class I source sample defined by 
\citet{KCH93}.   Each panel shows a $30'' \times 30''$ area.
The LRIS images of the five sources analyzed in this paper 
are given in Figure \ref{fig:iband}, and those sources we did not analyze 
are shown here.  Some of these objects are not detected in our observations,
which may indicate different source luminosities or circumstellar dust
geometries.
\label{fig:lris_all}}
\end{figure}

As discussed in \S \ref{sec:sample}, we obtained 0.9 $\mu$m images 
using Keck/LRIS of the entire sample
studied by \citet{KCH93}.  The details of these observations
are described in \S \ref{sec:lris}.  Although the analysis presented
above concentrated on only those sources that satisfied our selection
criteria (\S \ref{sec:sample}), we display the LRIS images for the larger 
sample in Figure \ref{fig:lris_all} for completeness.  

Several of these objects are detected strongly at 0.9 $\mu$m:
IRAS 04181+2654A, IRAS 04248+2612, 
IRAS 04263+2436, IRAS 04264+2433AB,
IRAS 04287+1802, IRAS 04303+2247, IRAS 04361+2547AB, IRAS 04368+2557,
and IRAS 04489+3402.  Four other sources exhibit faint smudges that may 
correspond to actual emission:  IRAS 04169+2702, IRAS 04181+2655, 
IRAS 04325+2402, and IRAS 04365+2635.  The remaining
sources were not detected, with a limiting $I$-band magnitude of $\sim 24$.
Photometry for these objects, 
measured within a $6\rlap{.}''3$ diameter aperture, is listed in 
\citet{WH04}.

Because these images are not very deep ($\le 300$s), 
we can not comment on whether the pattern of detections and
non-detections corresponds to variations in the underlying protostellar
luminosities, different inclinations, or other factors.  However, it is
clear from Figure \ref{fig:lris_all} that there are several sources
exhibiting bright scattered light structures; these sources are ideal
candidates for future work.

\epsscale{1.0}

\epsscale{1.0}

\epsscale{1.0}

\epsscale{1.0}

\epsscale{1.0}

\epsscale{1.0}

\epsscale{1.0}

\clearpage
\begin{deluxetable}{lcccccc}
\tabletypesize{\footnotesize}
\tablewidth{0pt}
\tablecaption{Spectral energy distributions for our sample \label{tab:seds}}
\tablehead{\colhead{$\lambda$ ($\mu$m)} & \multicolumn{5}{c}{Source Fluxes, 
$F_\nu$ (Jy)} & \colhead{reference$^a$} \\
&  04016+2610 &  04108+2803 &  04239+2436 &  04295+2251
&  04381+2540}
\startdata
0.55 & 2.42e-05 &   &   &   &   &  \citet{KCH93} \\
0.63 & 6.17e-05 &   &   &   &   &  \citet{KCH93} \\
0.79 & 0.0001 &   &   &   &   &  \citet{KCH93} \\
0.89 & 0.0001 & 3.23e-06 & 3.89e-05 & 2.23e-05 & 8.13e-06 &  This work \\
1.10 & 0.0003 &   &   &   &   &  \citet{PADGETT+99} \\
1.22 & 0.0055 & 0.0004 & 0.0010 & 0.0055 & 0.0005 &  \citet{KCH93} \\
1.22 &   &   &   & 0.0038 &   &  \citet{WKG97} \\
1.60 & 0.0095 &   &   &   &   &  \citet{PADGETT+99} \\
1.63 & 0.0217 & 0.0050 & 0.0075 & 0.0360 & 0.0018 &  \citet{KCH93} \\
1.63 & 0.0233 & 0.0242 & 0.0076 & 0.0229 & 0.0029 &  \citet{PK02} \\
1.63 & 0.0263 & 0.0043 & 0.0066 & 0.0220 &   &  \citet{WKG97} \\
1.87 & 0.0214 &   &   &   &   &  \citet{PADGETT+99} \\
2.05 & 0.0416 &   &   &   &   &  \citet{PADGETT+99} \\
2.19 & 0.1160 & 0.0226 & 0.0421 & 0.0943 & 0.0101 &  \citet{KCH93} \\
2.19 & 0.1045 & 0.0420 & 0.0464 & 0.0690 & 0.0147 &  \citet{PK02} \\
2.19 & 0.1055 & 0.0148 & 0.0439 & 0.0671 & 0.0149 &  \citet{WKG97} \\
3.45 & 0.4593 & 0.0492 & 0.2302 & 0.1667 & 0.0470 &  \citet{KCH93} \\
4.75 & 2.0465 & 0.1702 &   &   &   &  \citet{KCH93} \\
7.80 & 2.8 & 0.31 &   &   &   &  \citet{MYERS+87} \\
8.69 & 2.10 & 0.50 &   &   &   &  \citet{MYERS+87} \\
9.50 & 1.70 & 0.49 &   &   &   &  \citet{MYERS+87} \\
10.3 & 2.50 & 0.57 &   &   &   &  \citet{MYERS+87} \\
10.6 & 2.5682 & 0.6020 &   & 0.5883 & 0.2087 &  \citet{KCH93} \\
10.7 & 2.97 & 0.76 & 1.71 & 0.72 & 0.23 & \citet{KESSLER-SILACCI+05}
\\
11.6 & 3.40 & 0.44 &   &   &   &  \citet{MYERS+87} \\
12.0 & 3.8327 & 0.7647 & 1.7120 & 0.6215 & 0.4106 &  \citet{KCH93} \\
12.5 & 4.90 & 1.30 &   &   &   &  \citet{MYERS+87} \\
17.9 & 4.57 &   &   & 1.07 &   &  This work \\
21.0 & 10.6303 & 2.6094 & 4.4314 &   & 0.1979 &  \citet{KCH93} \\
25.0 & 16.3030 & 3.6498 & 6.9545 & 1.8718 & 2.7686 &  \citet{KCH93} \\
60.0 & 48.1371 & 7.6292 & 15.2222 & 3.4872 & 10.5312 &  \citet{KCH93} \\
100 & 62.2772 & 10.8225 & 17.1525 & 8.2097 & 14.9392 &  \citet{KCH93} \\
100 & 55.5046 &   &   &   &   &  \citet{KCH93} \\
160 & 46.6067 &   &   &   &   &  \citet{KCH93} \\
350 & 4.7686 &   &   &   &   &  \citet{KCH93} \\
450 & 3.2926 & 1.8094 &   &   &   &  \citet{KCH93} \\
450 & 2.8199 &   &   &   & 1.2000 &  \citet{HS00} \\
450 & 4.23 & 1.13 & 2.1400 & 2.6600 & 2.8199 &  \citet{YOUNG+03} \\
800 & 0.582 & 0.085 & 0.3330 & 0.2410 & 0.2890 &  \citet{MORIARTY-SCHIEVEN+94} \\
800 & 0.3143 & 0.8859 & 2.3303 &   &   &  \citet{KCH93} \\
850 & 0.5899 & 0.1700 & 0.4199 & 0.4199 & 0.5600 &  \citet{YOUNG+03} \\
850 & 0.43 &   &   &   & 0.24 &  \citet{HS00} \\
1100 & 0.180 & $<$0.100 & 0.114 & 0.094 & 0.116 &  \citet{MORIARTY-SCHIEVEN+94} \\
1100 & 0.5317 & 0.0882 & 0.0768 &   &   &  \citet{KCH93} \\
1300 & 0.150 & 0.039 & 0.170 & 0.115 & 0.300 &  \citet{MA01} \\
1300 & 0.0658 & 0.0271 & 0.0329 & 0.0484 & 0.0135 &  This work \\
2700 & $<$0.0068 &   &   &   & $<$0.0075 &  \citet{HOGERHEIJDE+97} \\
3000 & $<$0.0096 & $<$0.0210 & $<$0.0120 & $<$0.0210 &   &  \citet{OHASHI+96} \\
3000 &   & 0.0058 & 0.0120 & 0.0087 &   &  This work \\
3369 & $<$0.0043 &   &   &   & $<$0.0050 &  \citet{HOGERHEIJDE+97} \\
3440 & $<$0.0180 &   & $<$0.0240 &   &   &  \citet{SAITO+01} \\
20000 & 0.0005 &   &   &   &   &  \citet{LUCAS+00} \\
35000 & 0.0003 &   & 0.0001 &   &   &  \citet{LUCAS+00} \\
60000 & $<$0.0002 &   &   &   &   &  \citet{LUCAS+00} \\
\enddata
\tablerefs{$^a$--\citet{KCH93} compiled SEDs from new data and the literature,
and this work contains references for much of this photometry.}
\tablecomments{Different observations were conducted with different angular
resolutions, which explains some of the photometric variations.  In addition,
all of the sources in our sample except IRAS 04295+2251 are known to be
variable at short wavelengths \citep[e.g.,][]{PK02}.}
\end{deluxetable}

\clearpage
\begin{deluxetable}{lcccccccccccc}
\rotate
\tabletypesize{\scriptsize}
\tablewidth{0pt}
\tablecaption{Best-fit Models 
\label{tab:bestfits}}
\tablehead{\colhead{Source} & \colhead{$\chi_{\rm r,tot}^2$} & 
\colhead{$\chi^2_{\rm r,SED}$} & \colhead{$\chi^2_{\rm r,0.9 \mu m}$} & 
\colhead{$\chi^2_{\rm r,1mm}$} & \colhead{$\dot{M}$} & \colhead{$R_{\rm c}$} & 
\colhead{$M_{\rm env}$} & \colhead{$M_{\rm disk}$} & 
\colhead{$h_0$} & \colhead{$R_{\rm out}$} &
\colhead{$i$} & \colhead{$L_{\rm central}$} \\
 &  &  &  &  & 
(M$_{\odot}$ yr$^{-1}$) & 
(AU) & (M$_{\odot}$) & (M$_{\odot}$) & 
(AU) & (AU) & ($^{\circ}$) & (L$_{\odot}$)}
\startdata
\multicolumn{11}{c}{Rotating Infalling Envelope Models} \\
\hline
IRAS 04016+2610 & 4.39 & 1.57 & 1.40 & 1.41 & $8 \times 10^{-6}$ & 100 & $0.07$ & $...$ & ... & 2000 & 36 & 3.8 \\
IRAS 04108+2803B & 5.25 & 2.90 & 1.08 & 1.26 & $6 \times 10^{-6}$ & 30 & $0.05$ & $...$ & ... & 2000 & 26 & 0.7 \\
IRAS 04239+2436 & 6.89 & 3.93 & 1.17 & 1.77 & $8 \times 10^{-6}$ & 100 & $0.07$ & $...$ & ... & 2000 & 42 & 2.3 \\
IRAS 04295+2251 & 13.07 & 10.05 & 1.21 & 1.79 & $9 \times 10^{-6}$ & 100 & $0.01$ & $...$ & ... & 500 & 34 & 1.1 \\
IRAS 04381+2540 & 4.25 & 1.93 & 1.01 & 1.30 & $1 \times 10^{-5}$ & 100 & $0.10$ & $...$ & ... & 2000 & 30 & 0.6 \\
\hline
\multicolumn{11}{c}{Flared Disk Models} \\
\hline
IRAS 04016+2610 & 14.30 & 9.68 & 3.35 & 1.24 & $...$ & ... & $...$ & $0.10$ & 25 & 2000 & 52 & 2.8 \\
IRAS 04108+2803B & 12.81 & 10.30 & 1.28 & 1.21 & $...$ & ... & $...$ & $0.10$ & 25 & 500 & 52 & 2.3 \\
IRAS 04239+2436 & 11.21 & 8.18 & 1.73 & 1.30 & $...$ & ... & $...$ & $1.00$ & 25 & 500 & 45 & 1.0 \\
IRAS 04295+2251 & 13.98 & 10.60 & 2.26 & 1.11 & $...$ & ... & $...$ & $1.00$ & 15 & 1000 & 55 & 0.9 \\
IRAS 04381+2540 & 9.70 & 7.56 & 1.03 & 1.10 & $...$ & ... & $...$ & $0.70$ & 15 & 2000 & 56 & 0.5 \\
\hline
\multicolumn{11}{c}{Disk+Envelope Models} \\
\hline
IRAS 04016+2610 & 4.12 & 1.35 & 1.51 & 1.25 & $6 \times 10^{-6}$ & 100 & $0.05$ & $0.01$ & 15 & 2000 & 37 & 4.7 \\
IRAS 04108+2803B & 3.68 & 1.40 & 1.04 & 1.22 & $5 \times 10^{-6}$ & 30 & $5 \times 10^{-3}$ & $0.50$ & 15 & 500 & 24 & 0.4 \\
IRAS 04239+2436 & 4.73 & 1.67 & 1.72 & 1.33 & $3 \times 10^{-6}$ & 30 & $0.01$ & $0.50$ & 15 & 1000 & 34 & 1.5 \\
IRAS 04295+2251 & 5.83 & 2.75 & 1.93 & 1.13 & $5 \times 10^{-6}$ & 30 & $5 \times 10^{-3}$ & $1.00$ & 15 & 500 & 22 & 0.6 \\
IRAS 04381+2540 & 3.47 & 1.34 & 1.04 & 1.08 & $9 \times 10^{-6}$ & 30 & $0.01$ & $1.00$ & 15 & 1000 & 34 & 0.6 \\
\hline
\multicolumn{11}{c}{Disk+Extinction Models} \\
\hline
IRAS 04016+2610 & 6.28 & 1.82 & 3.20 & 1.25 & $...$ & ... & A$_{\rm V}$=10 & $0.20$ & 25 & 2000 & 49 & 5.3 \\
IRAS 04108+2803B & 3.34 & 1.07 & 1.05 & 1.21 & $...$ & ... & A$_{\rm V}$=25 & $0.60$ & 25 & 500 & 17 & 0.6 \\
IRAS 04239+2436 & 5.01 & 2.44 & 1.21 & 1.35 & $...$ & ... & A$_{\rm V}$=20 & $0.20$ & 20 & 2000 & 50 & 5.4 \\
IRAS 04295+2251 & 4.42 & 1.08 & 2.24 & 1.09 & $...$ & ... & A$_{\rm V}$=20 & $1.00$ & 15 & 500 & 49 & 1.2 \\
IRAS 04381+2540 & 3.27 & 1.12 & 1.06 & 1.07 & $...$ & ... & A$_{\rm V}$=20 & $0.70$ & 15 & 500 & 56 & 1.0 \\
\enddata
\tablecomments{Best-fit models for different density distributions considered
in \S \ref{sec:model}.  $\chi^2_{\rm r,tot}$ is the sum of the reduced chi-squared
residuals between the data and model for the SED, 0.9 $\mu$m, and 1 mm images.  The individual reduced chi squared
values for each dataset are denoted by $\chi^2_{\rm r, SED}$, $\chi^2_{\rm r, 0.9 \mu m}$, and $\chi^2_{\rm r, 1mm}$.
In the case of the disk+extinction models, entries in
the $M_{\rm env}$ column correspond to foreground extinction.
Although the disk masses listed in this Table are the results of our 
model-fitting, we believe that these are likely over-estimated 
(\S \ref{sec:dmass}).}
\end{deluxetable}

\clearpage
\begin{deluxetable}{lcccc}
\tabletypesize{\footnotesize}
\tablewidth{0pt}
\tablecaption{Comparison of compact and large-scale millimeter emission
\label{tab:compact}}
\tablehead{\colhead{Source} & \colhead{$R_{\rm data}$} & 
  \colhead{$R_{\rm disk}$} & \colhead{$R_{\rm env}$} &  
  \colhead{$R_{\rm disk+env}$}}
\startdata
IRAS 04016+2610 & 0.43 & 0.17 & 0.06 & 0.15 \\
IRAS 04108+2803B & 0.70 & 0.60 & 0.08 & 0.57 \\
IRAS 04239+2436 & 0.19 & 0.53 & 0.06 & 0.29 \\
IRAS 04295+2251 & 0.42 & 0.28 & 0.26 & 0.35 \\
IRAS 04381+2540 & 0.05 & 0.12 & 0.05 & 0.32 \\
\enddata
\tablerefs{In this table, $R_{\rm data}$ is the ratio of emission observed in 
our compact OVRO beam to the emission observed in the larger-beam survey of
\citet{MA01}.  $R_{\rm disk}$, $R_{\rm env}$, and $R_{\rm disk+env}$ are the
ratios one would measure for pure disk, pure envelope, and disk+envelope
models. For disk+extinction models (\S \ref{sec:d+ex}), the distribution
of large-scale material is unconstrained, and thus we do not calculate
expected ratios for these models.}
\end{deluxetable}


\begin{thebibliography}{100}
\expandafter\ifx\csname natexlab\endcsname\relax\def\natexlab#1{#1}\fi

\bibitem[{{Adams} {et~al.}(1987){Adams}, {Lada}, \& {Shu}}]{ALS87}
{Adams}, F.~C., {Lada}, C.~J., \& {Shu}, F.~H. 1987, \apj, 312, 788

\bibitem[{{Andr\'{e}} \& {Montmerle}(1994)}]{AM94}
{Andr\'{e}}, P. \& {Montmerle}, T. 1994, \apj, 420, 837

\bibitem[{{Andr\'{e}} {et~al.}(1993){Andr\'{e}}, {Ward-Thompson}, \&
  {Barsony}}]{AWB93}
{Andr\'{e}}, P., {Ward-Thompson}, D., \& {Barsony}, M. 1993, \apj, 406, 122

\bibitem[{{B{\` e}land} {et~al.}(1988){B{\` e}land}, {Boulade}, \&
  {Davidge}}]{BBD88}
{B{\` e}land}, S., {Boulade}, O., \& {Davidge}, T. 1988, Bulletin d'information
  du telescope Canada-France-Hawaii, 19, 16

\bibitem[{{Beichman} {et~al.}(1986){Beichman}, {Myers}, {Emerson}, {Harris},
  {Mathieu}, {Benson}, \& {Jennings}}]{BEICHMAN+86}
{Beichman}, C.~A., {Myers}, P.~C., {Emerson}, J.~P., {Harris}, S., {Mathieu},
  R., {Benson}, P.~J., \& {Jennings}, R.~E. 1986, \apj, 307, 337

\bibitem[{{Bell} \& {Lin}(1994)}]{BL94}
{Bell}, K.~R. \& {Lin}, D.~N.~C. 1994, \apj, 427, 987

\bibitem[{{Benson} \& {Myers}(1989)}]{BM89}
{Benson}, P.~J. \& {Myers}, P.~C. 1989, \apjs, 71, 89

\bibitem[{{Brown} \& {Chandler}(1999)}]{BC99}
{Brown}, D.~W. \& {Chandler}, C.~J. 1999, \mnras, 303, 855

\bibitem[{{Butner} {et~al.}(1991){Butner}, {Evans}, {Lester}, {Levreault}, \&
  {Strom}}]{BUTNER+91}
{Butner}, H.~M., {Evans}, N.~J., {Lester}, D.~F., {Levreault}, R.~M., \&
  {Strom}, S.~E. 1991, \apj, 376, 636

\bibitem[{{Calvet} \& {Gullbring}(1998)}]{CG98}
{Calvet}, N. \& {Gullbring}, E. 1998, \apj, 509, 802

\bibitem[{{Cassen} \& {Moosman}(1981)}]{CM81}
{Cassen}, P. \& {Moosman}, A. 1981, Icarus, 48, 353

\bibitem[{{Chandler} \& {Richer}(2000)}]{CR00}
{Chandler}, C.~J. \& {Richer}, J.~S. 2000, \apj, 530, 851

\bibitem[{{Chandler} {et~al.}(1996){Chandler}, {Terebey}, {Barsony}, {Moore},
  \& {Gautier}}]{CHANDLER+96}
{Chandler}, C.~J., {Terebey}, S., {Barsony}, M., {Moore}, T.~J.~T., \&
  {Gautier}, T.~N. 1996, \apj, 471, 308

\bibitem[{{Chiang} \& {Goldreich}(1997)}]{CG97}
{Chiang}, E.~I. \& {Goldreich}, P. 1997, \apj, 490, 368

\bibitem[{{Chiang} \& {Goldreich}(1999)}]{CG99}
---. 1999, \apj, 519, 279

\bibitem[{{D'Alessio} {et~al.}(2001){D'Alessio}, {Calvet}, \&
  {Hartmann}}]{DCH01}
{D'Alessio}, P., {Calvet}, N., \& {Hartmann}, L. 2001, \apj, 553, 321

\bibitem[{{D'Alessio} {et~al.}(1999){D'Alessio}, {Calvet}, {Hartmann},
  {Lizano}, \& {Cant{\' o}}}]{DALESSIO+99}
{D'Alessio}, P., {Calvet}, N., {Hartmann}, L., {Lizano}, S., \& {Cant{\' o}},
  J. 1999, \apj, 527, 893

\bibitem[{{Draine} \& {Lee}(1984)}]{DL84}
{Draine}, B.~T. \& {Lee}, H.~M. 1984, \apj, 285, 89

\bibitem[{{Draine} \& {Malhotra}(1993)}]{DM93}
{Draine}, B.~T. \& {Malhotra}, S. 1993, \apj, 414, 632

\bibitem[{{Duch\^{e}ne} {et~al.}(2004){Duch\^{e}ne}, {Bouvier}, {Bontemps},
  {Andr\'{e}}, \& {Motte}}]{DUCHENE+04}
{Duch\^{e}ne}, G., {Bouvier}, J., {Bontemps}, S., {Andr\'{e}}, P., \& {Motte},
  F. 2004, \aap, 427, 651

\bibitem[{{Dullemond} {et~al.}(2001){Dullemond}, {Dominik}, \& {Natta}}]{DDN01}
{Dullemond}, C.~P., {Dominik}, C., \& {Natta}, A. 2001, \apj, 560, 957

\bibitem[{{Eisner} {et~al.}(2003){Eisner}, {Lane}, {Akeson}, {Hillenbrand}, \&
  {Sargent}}]{EISNER+03}
{Eisner}, J.~A., {Lane}, B.~F., {Akeson}, R.~L., {Hillenbrand}, L., \&
  {Sargent}, A. 2003, \apj, 588, 360

\bibitem[{{Eisner} {et~al.}(2004){Eisner}, {Lane}, {Hillenbrand}, {Akeson}, \&
  {Sargent}}]{EISNER+04}
{Eisner}, J.~A., {Lane}, B.~F., {Hillenbrand}, L., {Akeson}, R., \& {Sargent},
  A. 2004, \apj, 613, 1049

\bibitem[{{Galli} \& {Shu}(1993{\natexlab{a}})}]{GS93a}
{Galli}, D. \& {Shu}, F.~H. 1993{\natexlab{a}}, \apj, 417, 220

\bibitem[{{Galli} \& {Shu}(1993{\natexlab{b}})}]{GS93b}
---. 1993{\natexlab{b}}, \apj, 417, 243

\bibitem[{{Goldreich} \& {Ward}(1973)}]{GW73}
{Goldreich}, P. \& {Ward}, W.~R. 1973, \apj, 183, 1051

\bibitem[{{Gomez} {et~al.}(1997){Gomez}, {Whitney}, \& {Kenyon}}]{GWK97}
{Gomez}, M., {Whitney}, B.~A., \& {Kenyon}, S.~J. 1997, \aj, 114, 1138

\bibitem[{{Grady} {et~al.}(1999){Grady}, {Woodgate}, {Bruhweiler}, {Boggess},
  {Plait}, {Lindler}, {Clampin}, \& {Kalas}}]{GRADY+99}
{Grady}, C.~A., {Woodgate}, B., {Bruhweiler}, F.~C., {Boggess}, A., {Plait},
  P., {Lindler}, D.~J., {Clampin}, M., \& {Kalas}, P. 1999, \apjl, 523, L151

\bibitem[{{Gullbring} {et~al.}(2000){Gullbring}, {Calvet}, {Muzerolle}, \&
  {Hartmann}}]{GULLBRING+00}
{Gullbring}, E., {Calvet}, N., {Muzerolle}, J., \& {Hartmann}, L. 2000, \apj,
  544, 927

\bibitem[{{Gullbring} {et~al.}(1998){Gullbring}, {Hartmann}, {Briceno}, \&
  {Calvet}}]{GULLBRING+98}
{Gullbring}, E., {Hartmann}, L., {Briceno}, C., \& {Calvet}, N. 1998, \apj,
  492, 323

\bibitem[{{Hartmann}(1998)}]{HARTMANN98}
{Hartmann}, L. 1998, {Accretion processes in star formation} (Accretion
  processes in star formation / Lee Hartmann.~Cambridge, UK ; New York :
  Cambridge University Press, 1998.~(Cambridge astrophysics series ; 32) ISBN
  0521435072.)

\bibitem[{{Hartmann} \& {Kenyon}(1996)}]{HK96}
{Hartmann}, L. \& {Kenyon}, S.~J. 1996, \araa, 34, 207

\bibitem[{{Hogerheijde}(2001)}]{HOGERHEIJDE01}
{Hogerheijde}, M.~R. 2001, \apj, 553, 618

\bibitem[{{Hogerheijde} \& {Sandell}(2000)}]{HS00}
{Hogerheijde}, M.~R. \& {Sandell}, G. 2000, \apj, 534, 880

\bibitem[{{Hogerheijde} {et~al.}(1997){Hogerheijde}, {van Dishoeck}, {Blake},
  \& {van Langevelde}}]{HOGERHEIJDE+97}
{Hogerheijde}, M.~R., {van Dishoeck}, E.~F., {Blake}, G.~A., \& {van
  Langevelde}, H.~J. 1997, \apj, 489, 293

\bibitem[{{Hogerheijde} {et~al.}(1998){Hogerheijde}, {van Dishoeck}, {Blake},
  \& {van Langevelde}}]{HOGERHEIJDE+98}
---. 1998, \apj, 502, 315

\bibitem[{{Ishii} {et~al.}(2004){Ishii}, {Tamura}, \& {Itoh}}]{ITI04}
{Ishii}, M., {Tamura}, M., \& {Itoh}, Y. 2004, \apj, 612, 956

\bibitem[{{Jones} \& {Puetter}(1993)}]{JP93}
{Jones}, B. \& {Puetter}, R.~C. 1993, in Proc. SPIE Vol. 1946, p. 610-621,
  Infrared Detectors and Instrumentation, Albert M. Fowler; Ed., 610--621

\bibitem[{{Keene} \& {Masson}(1990)}]{KM90}
{Keene}, J. \& {Masson}, C.~R. 1990, \apj, 355, 635

\bibitem[{{Kenyon} {et~al.}(1993{\natexlab{a}}){Kenyon}, {Calvet}, \&
  {Hartmann}}]{KCH93}
{Kenyon}, S.~J., {Calvet}, N., \& {Hartmann}, L. 1993{\natexlab{a}}, \apj, 414,
  676

\bibitem[{{Kenyon} \& {Hartmann}(1987)}]{KH87}
{Kenyon}, S.~J. \& {Hartmann}, L. 1987, \apj, 323, 714

\bibitem[{{Kenyon} \& {Hartmann}(1995)}]{KH95}
---. 1995, \apjs, 101, 117

\bibitem[{{Kenyon} {et~al.}(1990){Kenyon}, {Hartmann}, {Strom}, \&
  {Strom}}]{KENYON+90}
{Kenyon}, S.~J., {Hartmann}, L.~W., {Strom}, K.~M., \& {Strom}, S.~E. 1990,
  \aj, 99, 869

\bibitem[{{Kenyon} {et~al.}(1993{\natexlab{b}}){Kenyon}, {Whitney}, {Gomez}, \&
  {Hartmann}}]{KENYON+93}
{Kenyon}, S.~J., {Whitney}, B.~A., {Gomez}, M., \& {Hartmann}, L.
  1993{\natexlab{b}}, \apj, 414, 773

\bibitem[{{Kessler-Silacci} {et~al.}(2005){Kessler-Silacci}, {Hillenbrand},
  {Blake}, \& {Meyer}}]{KESSLER-SILACCI+05}
{Kessler-Silacci}, J.~E., {Hillenbrand}, L.~A., {Blake}, G.~A., \& {Meyer},
  M.~R. 2005, \apj, 622, 404

\bibitem[{{Koerner} \& {Sargent}(1995)}]{KS95}
{Koerner}, D.~W. \& {Sargent}, A.~I. 1995, \aj, 109, 2138

\bibitem[{{Lada}(1987)}]{LADA87}
{Lada}, C.~J. 1987, in IAU Symp. 115, Star Forming Regions, ed. M. Peimbert \&
  J. Jugaku (Dordrecht: Reidel), 1

\bibitem[{{Lada} \& {Wilking}(1984)}]{LW84}
{Lada}, C.~J. \& {Wilking}, B.~A. 1984, \apj, 287, 610

\bibitem[{{Ladd} {et~al.}(1995){Ladd}, {Fuller}, {Padman}, {Myers}, \&
  {Adams}}]{LADD+95}
{Ladd}, E.~F., {Fuller}, G.~A., {Padman}, R., {Myers}, P.~C., \& {Adams}, F.~C.
  1995, \apj, 439, 771

\bibitem[{{Landolt}(1992)}]{LANDOLT92}
{Landolt}, A.~U. 1992, \aj, 104, 340

\bibitem[{{Larson}(1969)}]{LARSON69}
{Larson}, R.~B. 1969, \mnras, 145, 271

\bibitem[{{Laughlin} \& {Bodenheimer}(1994)}]{LB94}
{Laughlin}, G. \& {Bodenheimer}, P. 1994, \apj, 436, 335

\bibitem[{{Lay} {et~al.}(1994){Lay}, {Carlstrom}, {Hills}, \&
  {Phillips}}]{LAY+94}
{Lay}, O.~P., {Carlstrom}, J.~E., {Hills}, R.~E., \& {Phillips}, T.~G. 1994,
  \apjl, 434, L75

\bibitem[{{Leinert} {et~al.}(2004){Leinert}, {van Boekel}, {Waters},
  {Chesneau}, {Malbet}, {K{\" o}hler}, {Jaffe}, {Ratzka}, {Dutrey},
  {Preibisch}, {Graser}, {Bakker}, {Chagnon}, {Cotton}, {Dominik}, {Dullemond},
  {Glazenborg-Kluttig}, {Glindemann}, {Henning}, {Hofmann}, {de Jong},
  {Lenzen}, {Ligori}, {Lopez}, {Meisner}, {Morel}, {Paresce}, {Pel},
  {Percheron}, {Perrin}, {Przygodda}, {Richichi}, {Sch{\" o}ller}, {Schuller},
  {Stecklum}, {van den Ancker}, {von der L{\" u}he}, \& {Weigelt}}]{LEINERT+04}
{Leinert}, C., {van Boekel}, R., {Waters}, L.~B.~F.~M., {Chesneau}, O.,
  {Malbet}, F., {K{\" o}hler}, R., {Jaffe}, W., {Ratzka}, T., {Dutrey}, A.,
  {Preibisch}, T., {Graser}, U., {Bakker}, E., {Chagnon}, G., {Cotton}, W.~D.,
  {Dominik}, C., {Dullemond}, C.~P., {Glazenborg-Kluttig}, A.~W., {Glindemann},
  A., {Henning}, T., {Hofmann}, K.-H., {de Jong}, J., {Lenzen}, R., {Ligori},
  S., {Lopez}, B., {Meisner}, J., {Morel}, S., {Paresce}, F., {Pel}, J.-W.,
  {Percheron}, I., {Perrin}, G., {Przygodda}, F., {Richichi}, A., {Sch{\"
  o}ller}, M., {Schuller}, P., {Stecklum}, B., {van den Ancker}, M.~E., {von
  der L{\" u}he}, O., \& {Weigelt}, G. 2004, \aap, 423, 537

\bibitem[{{Lucas} {et~al.}(2000){Lucas}, {Blundell}, \& {Roche}}]{LUCAS+00}
{Lucas}, P.~W., {Blundell}, K.~M., \& {Roche}, P.~F. 2000, \mnras, 318, 526

\bibitem[{{Lucas} \& {Roche}(1998)}]{LR98}
{Lucas}, P.~W. \& {Roche}, P.~F. 1998, \mnras, 299, 699

\bibitem[{{Mathis} {et~al.}(1977){Mathis}, {Rumpl}, \& {Nordsieck}}]{MRN77}
{Mathis}, J.~S., {Rumpl}, W., \& {Nordsieck}, K.~H. 1977, \apj, 217, 425

\bibitem[{{McCaughrean} \& {O'Dell}(1996)}]{MO96}
{McCaughrean}, M.~J. \& {O'Dell}, C.~R. 1996, \aj, 111, 1977

\bibitem[{{Monnier} {et~al.}(2004){Monnier}, {Tuthill}, {Ireland}, {Cohen}, \&
  {Tannirkulam}}]{MONNIER+04}
{Monnier}, J.~D., {Tuthill}, P.~G., {Ireland}, M.~J., {Cohen}, R., \&
  {Tannirkulam}, A. 2004, American Astronomical Society Meeting Abstracts, 205,

\bibitem[{{Moriarty-Schieven} {et~al.}(1994){Moriarty-Schieven}, {Wannier},
  {Keene}, \& {Tamura}}]{MORIARTY-SCHIEVEN+94}
{Moriarty-Schieven}, G.~H., {Wannier}, P.~G., {Keene}, J., \& {Tamura}, M.
  1994, \apj, 436, 800

\bibitem[{{Moriarty-Schieven} {et~al.}(1992){Moriarty-Schieven}, {Wannier},
  {Tamura}, \& {Keene}}]{MORIARTY-SCHIEVEN+92}
{Moriarty-Schieven}, G.~H., {Wannier}, P.~G., {Tamura}, M., \& {Keene}, J.
  1992, \apj, 400, 260

\bibitem[{{Motte} \& {Andr{\' e}}(2001)}]{MA01}
{Motte}, F. \& {Andr{\' e}}, P. 2001, \aap, 365, 440

\bibitem[{{Myers} {et~al.}(1987){Myers}, {Fuller}, {Mathieu}, {Beichman},
  {Benson}, {Schild}, \& {Emerson}}]{MYERS+87}
{Myers}, P.~C., {Fuller}, G.~A., {Mathieu}, R.~D., {Beichman}, C.~A., {Benson},
  P.~J., {Schild}, R.~E., \& {Emerson}, J.~P. 1987, \apj, 319, 340

\bibitem[{{Myers} {et~al.}(1988){Myers}, {Heyer}, {Snell}, \&
  {Goldsmith}}]{MYERS+88}
{Myers}, P.~C., {Heyer}, M., {Snell}, R.~L., \& {Goldsmith}, P.~F. 1988, \apj,
  324, 907

\bibitem[{{Nakazato} {et~al.}(2003){Nakazato}, {Nakamoto}, \&
  {Umemura}}]{NNU03}
{Nakazato}, T., {Nakamoto}, T., \& {Umemura}, M. 2003, \apj, 583, 322

\bibitem[{{Ohashi} {et~al.}(1996){Ohashi}, {Hayashi}, {Kawabe}, \&
  {Ishiguro}}]{OHASHI+96}
{Ohashi}, N., {Hayashi}, M., {Kawabe}, R., \& {Ishiguro}, M. 1996, \apj, 466,
  317

\bibitem[{{Oke} {et~al.}(1995){Oke}, {Cohen}, {Carr}, {Cromer}, {Dingizian},
  {Harris}, {Labrecque}, {Lucinio}, {Schaal}, {Epps}, \& {Miller}}]{OKE+95}
{Oke}, J.~B., {Cohen}, J.~G., {Carr}, M., {Cromer}, J., {Dingizian}, A.,
  {Harris}, F.~H., {Labrecque}, S., {Lucinio}, R., {Schaal}, W., {Epps}, H., \&
  {Miller}, J. 1995, \pasp, 107, 375

\bibitem[{{Osorio} {et~al.}(2003){Osorio}, {D'Alessio}, {Muzerolle}, {Calvet},
  \& {Hartmann}}]{OSORIO+03}
{Osorio}, M., {D'Alessio}, P., {Muzerolle}, J., {Calvet}, N., \& {Hartmann}, L.
  2003, \apj, 586, 1148

\bibitem[{{Padgett} {et~al.}(1999){Padgett}, {Brandner}, {Stapelfeldt},
  {Strom}, {Terebey}, \& {Koerner}}]{PADGETT+99}
{Padgett}, D.~L., {Brandner}, W., {Stapelfeldt}, K.~R., {Strom}, S.~E.,
  {Terebey}, S., \& {Koerner}, D. 1999, \aj, 117, 1490

\bibitem[{{Park} \& {Kenyon}(2002)}]{PK02}
{Park}, S. \& {Kenyon}, S.~J. 2002, \aj, 123, 3370


\bibitem[Pascucci et al.(2004)]{PASCUCCI+04} Pascucci, I., Wolf, 
S., Steinacker, J., Dullemond, C.~P., Henning, T., Niccolini, G., Woitke, 
P., \& Lopez, B.\ 2004, \aap, 417, 793 

\bibitem[{{Rodriguez} {et~al.}(1998){Rodriguez}, {D'Alessio}, {Wilner}, {Ho},
  {Torrelles}, {Curiel}, {Gomez}, {Lizano}, {Pedlar}, {Canto}, \&
  {Raga}}]{RODRIGUEZ+98}
{Rodriguez}, L.~F., {D'Alessio}, P., {Wilner}, D.~J., {Ho}, P.~T.~P.,
  {Torrelles}, J.~M., {Curiel}, S., {Gomez}, Y., {Lizano}, S., {Pedlar}, A.,
  {Canto}, J., \& {Raga}, A.~C. 1998, \nat, 395, 355

\bibitem[{{Saito} {et~al.}(2001){Saito}, {Kawabe}, {Kitamura}, \&
  {Sunada}}]{SAITO+01}
{Saito}, M., {Kawabe}, R., {Kitamura}, Y., \& {Sunada}, K. 2001, \apj, 547, 840

\bibitem[{{Sault} {et~al.}(1995){Sault}, {Teuben}, \& {Wright}}]{STW95}
{Sault}, R.~J., {Teuben}, P.~J., \& {Wright}, M.~C.~H. 1995, in ASP Conf. Ser.
  77: Astronomical Data Analysis Software and Systems IV, 433--+

\bibitem[{{Scoville} {et~al.}(1993){Scoville}, {Carlstrom}, {Chandler},
  {Phillips}, {Scott}, {Tilanus}, \& {Wang}}]{SCOVILLE+93}
{Scoville}, N.~Z., {Carlstrom}, J.~E., {Chandler}, C.~J., {Phillips}, J.~A.,
  {Scott}, S.~L., {Tilanus}, R.~P.~J., \& {Wang}, Z. 1993, \pasp, 105, 1482

\bibitem[{{Semenov} {et~al.}(2004){Semenov}, {Pavlyuchenkov}, {Schreyer},
  {Henning}, {Dullemond}, \& {Bacmann}}]{SEMENOV+04}
{Semenov}, D., {Pavlyuchenkov}, Y., {Schreyer}, K., {Henning}, T., {Dullemond},
  C., \& {Bacmann}, A. 2004, ArXiv Astrophysics e-prints

\bibitem[{{Shakura} \& {Sunyaev}(1973)}]{SS73}
{Shakura}, N.~I. \& {Sunyaev}, R.~A. 1973, \aap, 24, 337

\bibitem[{{Shu} {et~al.}(1993){Shu}, {Najita}, {Galli}, {Ostriker}, \&
  {Lizano}}]{SHU+93}
{Shu}, F., {Najita}, J., {Galli}, D., {Ostriker}, E., \& {Lizano}, S. 1993, in
  Protostars and Planets III, 3--45

\bibitem[{{Shu}(1977)}]{SHU77}
{Shu}, F.~H. 1977, \apj, 214, 488

\bibitem[{{Shu} {et~al.}(1987){Shu}, {Adams}, \& {Lizano}}]{SAL87}
{Shu}, F.~H., {Adams}, F.~C., \& {Lizano}, S. 1987, \araa, 25, 23

\bibitem[{{Strom} {et~al.}(1976){Strom}, {Strom}, \& {Vrba}}]{SSV76}
{Strom}, K.~M., {Strom}, S.~E., \& {Vrba}, F.~J. 1976, \aj, 81, 320

\bibitem[{{Su} {et~al.}(2005){Su}, {Rieke}, {Misselt}, {Stansberry},
  {Moro-Martin}, {Stapelfeldt}, {Werner}, {Trilling}, {Bendo}, {Gordon},
  {Hines}, {Wyatt}, {Holland}, {Marengo}, {Megeath}, \& {Fazio}}]{SU+05}
{Su}, K.~Y.~L., {Rieke}, G.~H., {Misselt}, K.~A., {Stansberry}, J.~A.,
  {Moro-Martin}, A., {Stapelfeldt}, K.~R., {Werner}, M.~W., {Trilling}, D.~E.,
  {Bendo}, G.~J., {Gordon}, K.~D., {Hines}, D.~C., {Wyatt}, M.~C., {Holland},
  W.~S., {Marengo}, M., {Megeath}, S.~T., \& {Fazio}, G.~G. 2005, ArXiv
  Astrophysics e-prints

\bibitem[{{Tamura} {et~al.}(1991){Tamura}, {Gatley}, {Waller}, \&
  {Werner}}]{TAMURA+91}
{Tamura}, M., {Gatley}, I., {Waller}, W., \& {Werner}, M.~W. 1991, \apjl, 374,
  L25

\bibitem[{{Terebey} {et~al.}(1984){Terebey}, {Shu}, \& {Cassen}}]{TSC84}
{Terebey}, S., {Shu}, F.~H., \& {Cassen}, P. 1984, \apj, 286, 529

\bibitem[{{Terebey} {et~al.}(1989){Terebey}, {Vogel}, \& {Myers}}]{TVM89}
{Terebey}, S., {Vogel}, S.~N., \& {Myers}, P.~C. 1989, \apj, 340, 472

\bibitem[{{Terquem} \& {Bertout}(1996)}]{TB96}
{Terquem}, C. \& {Bertout}, C. 1996, \mnras, 279, 415

\bibitem[{{Ulrich}(1976)}]{ULRICH76}
{Ulrich}, R.~K. 1976, \apj, 210, 377

\bibitem[{{Watson} {et~al.}(2004){Watson}, {Kemper}, {Calvet}, {Keller},
  {Hartmann}, {Forrest}, {Chen}, {Uchida}, {Green}, {Sargent}, {Sloan},
  {Herter}, {Brandl}, {Houck}, {Najita}, {D'Alessio}, {Myers}, {Barry}, {Hall},
  \& {Morris}}]{WATSON+04}
{Watson}, D.~M., {Kemper}, F., {Calvet}, N., {Keller}, L.~D.~{Furlan}, E.,
  {Hartmann}, L., {Forrest}, W.~J., {Chen}, C.~H., {Uchida}, K.~I., {Green},
  J.~D., {Sargent}, B., {Sloan}, G.~C., {Herter}, T.~L., {Brandl}, B.~R.,
  {Houck}, J.~R., {Najita}, J., {D'Alessio}, P., {Myers}, P.~C., {Barry},
  D.~J., {Hall}, P., \& {Morris}, P.~W. 2004, \apjs, 154, 391

\bibitem[{{Weingartner} \& {Draine}(2001)}]{WD01}
{Weingartner}, J.~C. \& {Draine}, B.~T. 2001, \apj, 548, 296

\bibitem[{{White} {et~al.}(2000){White}, {Liseau}, {Men'shchikov},
  {Justtanont}, {Nisini}, {Benedettini}, {Caux}, {Ceccarelli}, {Correia},
  {Giannini}, {Kaufman}, {Lorenzetti}, {Molinari}, {Saraceno}, {Smith},
  {Spinoglio}, \& {Tommasi}}]{WHITE+00}
{White}, G.~J., {Liseau}, R., {Men'shchikov}, A.~B., {Justtanont}, K.,
  {Nisini}, B., {Benedettini}, M., {Caux}, E., {Ceccarelli}, C., {Correia},
  J.~C., {Giannini}, T., {Kaufman}, M., {Lorenzetti}, D., {Molinari}, S.,
  {Saraceno}, P., {Smith}, H.~A., {Spinoglio}, L., \& {Tommasi}, E. 2000, \aap,
  364, 741

\bibitem[{{White} \& {Hillenbrand}(2004)}]{WH04}
{White}, R.~J. \& {Hillenbrand}, L.~A. 2004, \apj, 616, 998

\bibitem[{{Whitney} \& {Hartmann}(1992)}]{WH92}
{Whitney}, B.~A. \& {Hartmann}, L. 1992, \apj, 395, 529

\bibitem[{{Whitney} {et~al.}(1997){Whitney}, {Kenyon}, \& {Gomez}}]{WKG97}
{Whitney}, B.~A., {Kenyon}, S.~J., \& {Gomez}, M. 1997, \apj, 485, 703

\bibitem[{{Whitney} {et~al.}(2003{\natexlab{a}}){Whitney}, {Wood}, {Bjorkman},
  \& {Cohen}}]{WHITNEY+03b}
{Whitney}, B.~A., {Wood}, K., {Bjorkman}, J.~E., \& {Cohen}, M.
  2003{\natexlab{a}}, \apj, 598, 1079

\bibitem[{{Whitney} {et~al.}(2003{\natexlab{b}}){Whitney}, {Wood}, {Bjorkman},
  \& {Wolff}}]{WHITNEY+03a}
{Whitney}, B.~A., {Wood}, K., {Bjorkman}, J.~E., \& {Wolff}, M.~J.
  2003{\natexlab{b}}, \apj, 591, 1049

\bibitem[{{Wolf}(2003)}]{WOLF03}
{Wolf}, S. 2003, Computer Physics Communications, 150, 99

\bibitem[{{Wolf} \& {Henning}(2000)}]{WH00}
{Wolf}, S. \& {Henning}, T. 2000, Computer Physics Communications, 132, 166

\bibitem[Wolf et al.(1999)]{WHS99} Wolf, S., Henning, T., \& 
Stecklum, B.\ 1999, \aap, 349, 839 

\bibitem[{{Wolf} {et~al.}(2003){Wolf}, {Padgett}, \& {Stapelfeldt}}]{WPS03}
{Wolf}, S., {Padgett}, D.~L., \& {Stapelfeldt}, K.~R. 2003, \apj, 588, 373

\bibitem[{{Wood} {et~al.}(2002){Wood}, {Wolff}, {Bjorkman}, \&
  {Whitney}}]{WOOD+02}
{Wood}, K., {Wolff}, M.~J., {Bjorkman}, J.~E., \& {Whitney}, B. 2002, \apj,
  564, 887

\bibitem[{{Yorke} {et~al.}(1995){Yorke}, {Bodenheimer}, \& {Laughlin}}]{YBL95}
{Yorke}, H.~W., {Bodenheimer}, P., \& {Laughlin}, G. 1995, \apj, 443, 199

\bibitem[{{Young} {et~al.}(2003){Young}, {Shirley}, {Evans}, \&
  {Rawlings}}]{YOUNG+03}
{Young}, C.~H., {Shirley}, Y.~L., {Evans}, N.~J., \& {Rawlings}, J.~M.~C. 2003,
  \apjs, 145, 111

\end{thebibliography}
\end{document}